\documentclass[9pt,numbers,sectionbib,sn-basic]{sn-jnl}

\usepackage{enumitem}
\usepackage{multirow}
\usepackage{hyperref}
\usepackage{verbatim}
\usepackage{amsmath,amstext,mathrsfs}
\usepackage[all]{hypcap} 
\usepackage{afterpage}    
\usepackage{marginnote}
\usepackage{makecell}
\usepackage{subfigure}
\usepackage{xfrac}

\usepackage{graphicx}
\usepackage{amssymb}		
\usepackage{latexsym}

\begin{document}

\jyear{2023}%

\theoremstyle{thmstyleone}%
\newtheorem{theorem}{Theorem}
\newtheorem{proposition}[theorem]{Proposition}%

\theoremstyle{thmstyletwo}%
\newtheorem{example}{Example}%
\newtheorem{remark}{Remark}%

\theoremstyle{thmstylethree}%
\newtheorem{definition}{Definition}%

\raggedbottom


\AtBeginEnvironment{subappendices}{%
\chapter*{Appendix}
\addcontentsline{toc}{chapter}{Appendices}
}

\AtEndEnvironment{subappendices}{%
}

\newcommand*{\kh}{} 
\newcommand*{\li}{}
\newcommand*{\avi}{}

\newcommand\aj{{AJ}}
\newcommand\prl{{PRL}}
\newcommand\araa{{ARA\&A}}
\newcommand\apj{{ApJ}}
\newcommand\apjl{{ApJL}}     
\newcommand\apjs{{ApJS}}
\newcommand\aap{{A\&A}}
\newcommand\aapr{{A\&A~Rv}}
\newcommand\aaps{{A\&AS}}
\newcommand\apss{{Astrophysics and Space Science}}
\newcommand\mnras{{MNRAS}}
\newcommand\ssr{{Space Science Review}}
\newcommand\jqsrt{{Journal of Quantitative Spectroscopy \& Radiative Transfer}}
\newcommand\pra{{Physical Review A }}
\newcommand\pre{{Physical Review E }}
\newcommand\nat{{Nat}}
\newcommand\jgr{{Journal of Geophysical Research }}
\newcommand\sovast{{Soviet Astronomy}}

\title[{\kh} Filaments in ISM]{{\kh Neutral Hydrogen} Filaments in interstellar media: Are they physical?}

\author*[1]{\fnm{Ka Ho} \sur{Yuen}}\email{kyuen@lanl.gov}
\author[1,2]{\fnm{Ka Wai} \sur{Ho}}
\author[3]{\fnm{Chi Yan} \sur{Law}}
\author[4]{\fnm{Avi} \sur{Chen}}

\affil*[1]{\orgdiv{Theoretical Division}, \orgname{Los Alamos National Laboratory}, \orgaddress{\city{Los Alamos}, \postcode{87545}, \state{NM}, \country{USA}}}
\affil[2]{\orgdiv{Department of Astronomy}, \orgname{University of Wisconsin-Madison}, \orgaddress{\street{454 N. Charter St}, \city{Madison}, \postcode{53715}, \state{WI}, \country{USA}}}
\affil[3]{\orgdiv{Department of Space, Earth \& Environment}, \orgname{Chalmers University of Technology}, \orgaddress{\postcode{SE-412 96},\city{Gothenburg}, \country{Sweden}}}
\affil[4]{\orgdiv{ Department of Physics and Astronomy}, \orgname{Rutgers University}, \orgaddress{\street{136 Frelinghuysen Rd}, \city{Piscataway}, \postcode{08854}, \state{NJ}, \country{USA}}}

\abstract{
The trending term "filament" is extensively used in the interstellar medium (ISM) and the star formation community, and is believed to be one of the most important objects that gauge molecular cloud and star formation. However, the physical definition of these ubiquitous, elongated, high contrast features is poorly defined and still actively debated. Despite the absence of a unified consensus, filaments are believed to be involved in many important physical processes from galaxy structure formation to the emergence of protostellar objects. Therefore, understanding how filaments form, what constrains their growth, and their general physical properties, are extremely important for theorists and observers who study the dynamics of the ISM and consequent star formations. This review serves as a collection of the community's views and develops the concept of "filaments" in the context of the ISM and star-forming clouds. Observationally, filaments are seen across the entire sky and often carry an aspect ratio of the order of hundreds. In the context of the ISM, filaments are believed to form by stretching and tearing from magnetized ISM turbulence. ISM filaments are subjected to heating and cooling phases, and are likely to be magnetically aligned. Cold clouds are formed inside ISM due to turbulence instability. This review updates the understanding of ISM filaments in the community.}


\maketitle

\section{Overview: What are "filaments" in the context of ISM? }\label{sec1}

Since early astronomical observations of the 'dark lane' that stands out in the sky in contrast to the bright galactic background, which is opaque in visible to near-infrared wavelengths, filamentary structures have been presented across the different phases of the interstellar medium across a wide range of scales from kilo-parsec to sub-parsec scales (See \citealt{2023ARA&A..61...19M} for a review). These structures are elongated features that have an aspect ratio significantly larger than unity and are an essential component in our understanding of the formation and evolution of stars, star clusters and Galaxies. 

The atomic hydrogen cold neutral medium (HI-CNM) is a popular and important astrophysical observable. Observations like HI4PI \citep{2015A&A...578A..78K}, GALFA \citep{2018ApJS..234....2P}, THOR-HI \citep{2016ApJ...822...83B}, and FAST (e.g. \citealt{cattail}) have advanced our understanding of CNM, including its spatial distribution \citep{2018A&A...619A..58K}, its relation to magnetic fields \citep{2003ApJ...586.1067H,2015PhRvL.115x1302C}, and its connection to the underlying molecular phase \citep{2017NJPh...19f5003K}. CNM is ubiquitous in the interstellar medium (ISM), spanning from high latitude \citep{2015PhRvL.115x1302C} to the galactic plane \citep{Soler2020}, and is highly filamentary with filaments aligned along the magnetic field direction \citep{2015PhRvL.115x1302C,YL17a}. This feature of CNM makes it one of the most important B-field probes in modern astronomy, with vast applications from cosmology \citep{Clark19} to diffuse molecular cloud studies (e.g., \citealt{LY18a}). The nature of CNM is also very important in understanding the formation of molecular clouds. With a density of $10cm^{-3}$ or higher the CNM is already near the critical density for H2 formation $(50 cm^{-3})$. Due to the rapid conversion between HI and H2 \citep{2016ApJ...822...83B}, the formed CNM can quickly convert into molecular hydrogen and experience Jeans instability, leading to the formation of molecular clouds.

While filamentary structures are expected to play a crucial role in structure formation and evolution in the ISM, many open questions remain. We list a few of these questions, which motivate and form the backbone of this review.

\subsection{Why are observed CNM filaments in the ISM only parallel to B-field while simulations contain both parallel and perpendicular?}

Our understanding of the physics of the multiphase is rather limited and hasn't advanced much from the description of the two-phase model of neutral media proposed in the late 20th century (\citealt{1977ApJ...218..148M}, left of Fig 2).
The ISM in our Milky Way galaxy is threaded by ubiquitous turbulent magnetic fields. This was realized long ago. For instance, in the 2004 ARA\&A review, \citeauthor{2004ARAnA..42..275S} (\citeyear{2004ARAnA..42..275S}) wrote: {\it "One of the most important developments in the field of interstellar gas ...was that ... most processes and structures are strongly affected by turbulence"}.

\begin{figure}
    \centering
    \includegraphics[width=5.5cm]{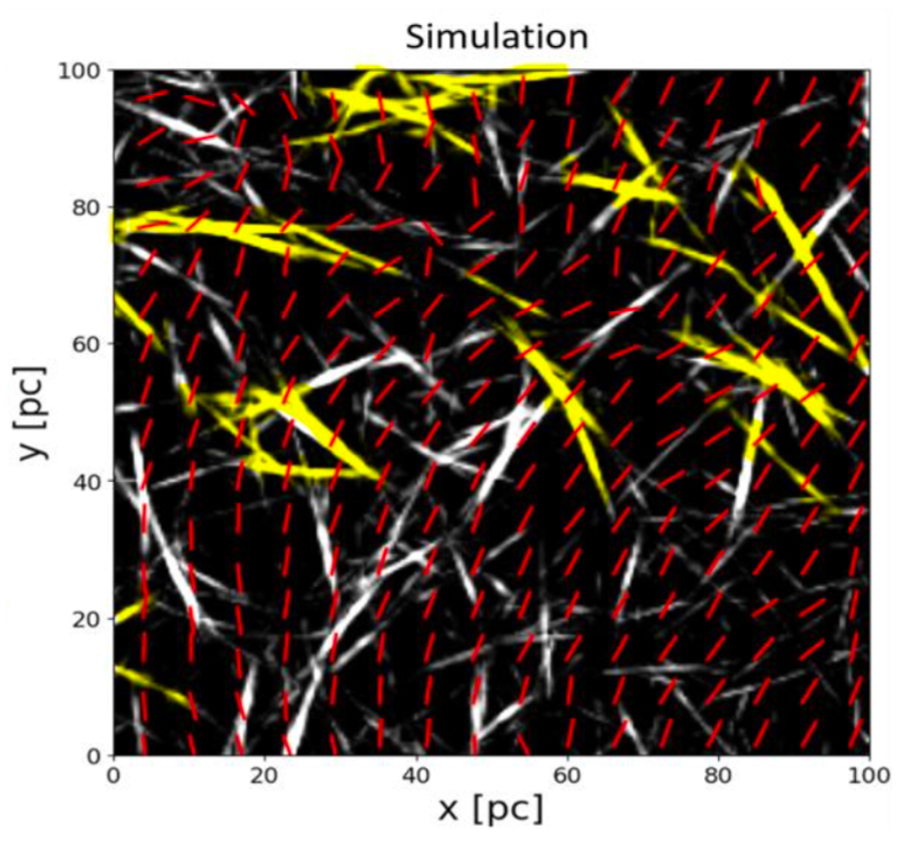}
    \caption{Extracted filaments using {\avi Rolling Hough Transform (RHT)} in multiphase numerical simulation from \cite{2017NJPh...19f5003K} (Credits: A. Kritsuk). The underlying vectors for both figures are the magnetic field directions in their retrospectives. We highlight the parallel filaments (filaments that are parallel to B-field) as white, the perpendicular filaments (filaments that are perpendicular to B-field) as yellow on the RHS to illustrate the differences between observations and simulations: There are almost no perpendicular filaments in observed HI data, but a vast number of filaments in simulations are perpendicular to B.}
    \label{fig:fig1}
\end{figure}

The ubiquity and linearity of CNM can best be explained with MHD turbulence theory. Filamentary features are mainly formed by: (1) low-density fluctuations along the magnetic field and (2) high-density shock compression perpendicular to the magnetic field. Isothermal numerical simulations \citep{2007ApJ...658..423K,2010ApJ...720..742K,2018ApJ...853..173K,LY18a} suggest that CNM filaments form both parallel and perpendicular to the magnetic field  (Fig. \ref{fig:fig1}). The number of parallel and perpendicular filaments have a comparable population in simulations, but it does not seem to be the case in observations (See, e.g. \cite{2015PhRvL.115x1302C,2022AnA...657A...1S}). 

Most HI linear features that are proposed candidates of CNM \citep{Clark19}, are preferentially parallel to the magnetic field (\citealt{2014ApJ...789...82C, 2015PhRvL.115x1302C}). Only in very rare circumstances do we observe CNM perpendicular to the magnetic field \citep{2017NJPh...19f5003K,2018ApJ...853..173K} . A major scientific question regarding CNM candidate is: why are filaments exclusively parallel to magnetic fields in observations and not in numerical simulations? (See Fig \ref{fig:fig1}.)

\subsection{Why is the aspect ratio of observed CNM in the hundreds while in simulations they are not?}
The observed parallel HI filaments generally have a very high aspect ratio, ranging from 30 to 100 (see Tab.~\ref{tab:filament_ar}) and are suggested to be associated with cold features \citep{Clark19}. Given the ubiquity of these long, cold, and thin filaments across the full sky ($\sim$ 0.1 pc thick, \citealt{2021A&A...654A..91K}), there must be a universal mechanism favoring the formation of such long CNM structures. One of the most common explanations is the equally ubiquitous magnetized turbulence, shaping the multiphase media according to the Goldreich-Sridhar scaling (hereafter GS95) ( \citealt{GS95}, see also \citealt{LV99,2019ApJ...878..157X}). This scaling predicts that filaments in isothermal MHD turbulence should scale as $l_\parallel/l_\perp\sim M_A^{-4/3} (L_{inj}/l_\perp )^{1/3}$. However, the observed filaments are much longer than this prediction when realistic values of the Alfvenic Mach number $M_A$, turbulence injection scale $L_{inj}$, and thickness $l_\perp$ of cold neutral media are considered (green line in Fig. 3). Another recently proposed mechanism favoring the formation of long filaments is small-scale reconnection \citep{2021A&A...654A..91K}. Yet no theoretical models explain how the statistics of the aspect ratio are related to the properties of small-scale reconnection.

Simulations \citep{instability} have shown that cold filaments formed in multiphase media are considerably shorter in length than what is observed (Fig \ref{fig:fig3}). Even for simulation with adaptive mesh refinement (e.g. \citealt{2022MNRAS.512.4765S}), the HI filaments are still much shorter than what we observe in HI maps. The aspect ratios in these simulations is also significantly shorter than what GS95 predicted (\citealt{instability}, Fig \ref{fig:fig3}), suggesting that the formation of long cold filaments may not be solely attributed to the stretching and tearing of magnetized turbulence. This raises the question of whether the GS95 scaling is an appropriate description for modeling the statistics of filament aspect ratios. It is known that the GS95 critical balance does not hold during the UNM stage. \cite{instability} suggests that the "cold filaments" identified through observations may not be physically stable and could be subject to a new instability, which naturally breaks down the filaments into clumps with shorter aspect ratios. This prompts us to inquire about the origin of the very long features observed in these observations.

\begin{figure}
    \centering
    \includegraphics[width=11cm]{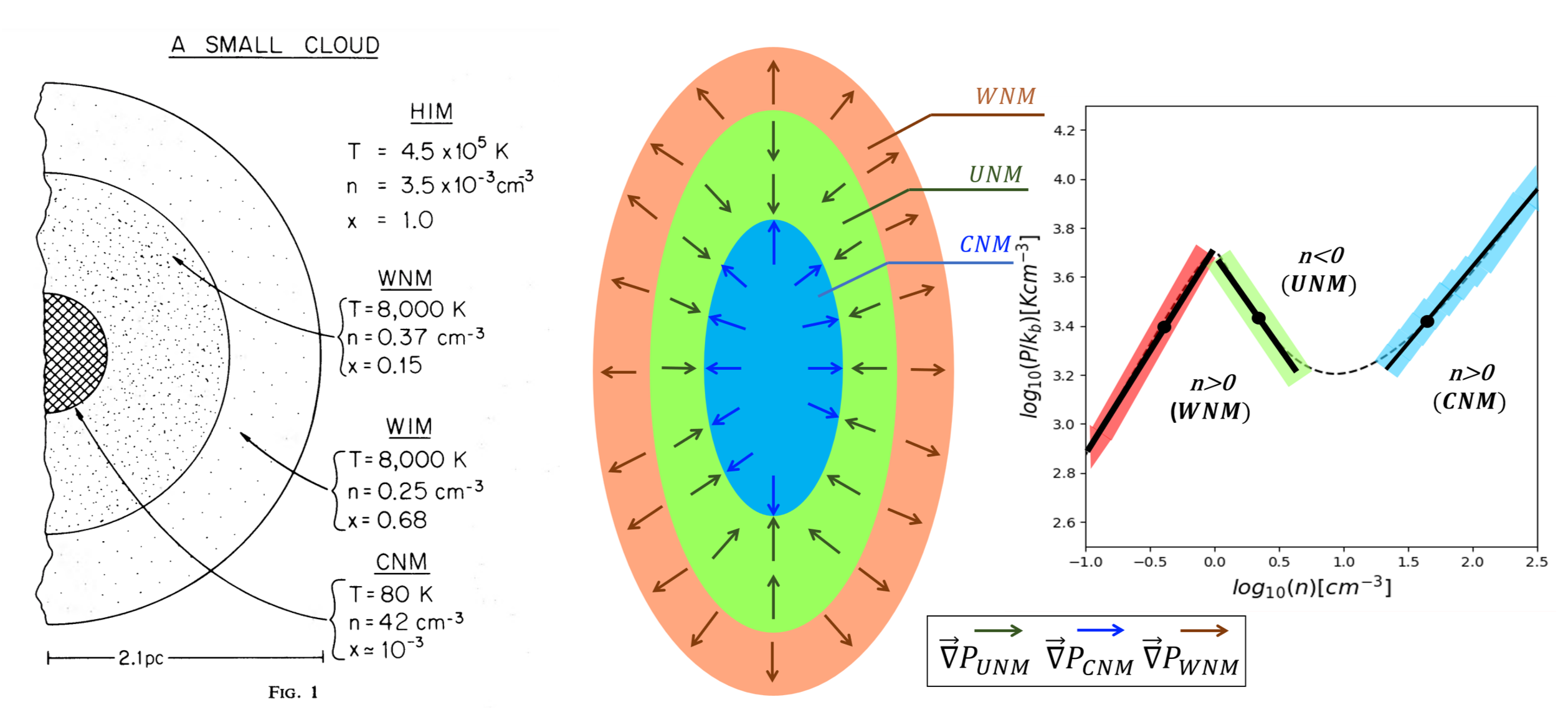}
    \caption{Illustration of the Multiphase model. Left: Model from \cite{1977ApJ...218..148M}. In their model, they only recognize two phases in the neutral stage, namely the warm neutral media (WNM) and cold neutral media (CNM). Middle and right panels: Modern understanding of the multiphase model and the respective phase diagram. From the phase diagram (right panel), we can identify three separate phases (middle panel). The new unstable phase, unstable neutral media (UNM), possesses a very special property: (1) it is a transient phase and in unstable equilibrium, i.e., the adiabatic index $n < 0$. (2) the pressure force exerted by UNM is attractive (green vectors in the middle panel). If we consider the pressure balances between the phases, the contracting pressure force from UNM (middle) is counterbalanced by the repelling force from the CNM. This force balance limits the length of the CNM \citep{instability}.}
    \label{fig:fig2}
\end{figure}

\begin{figure}
    \centering
    \includegraphics[width=7.5cm]{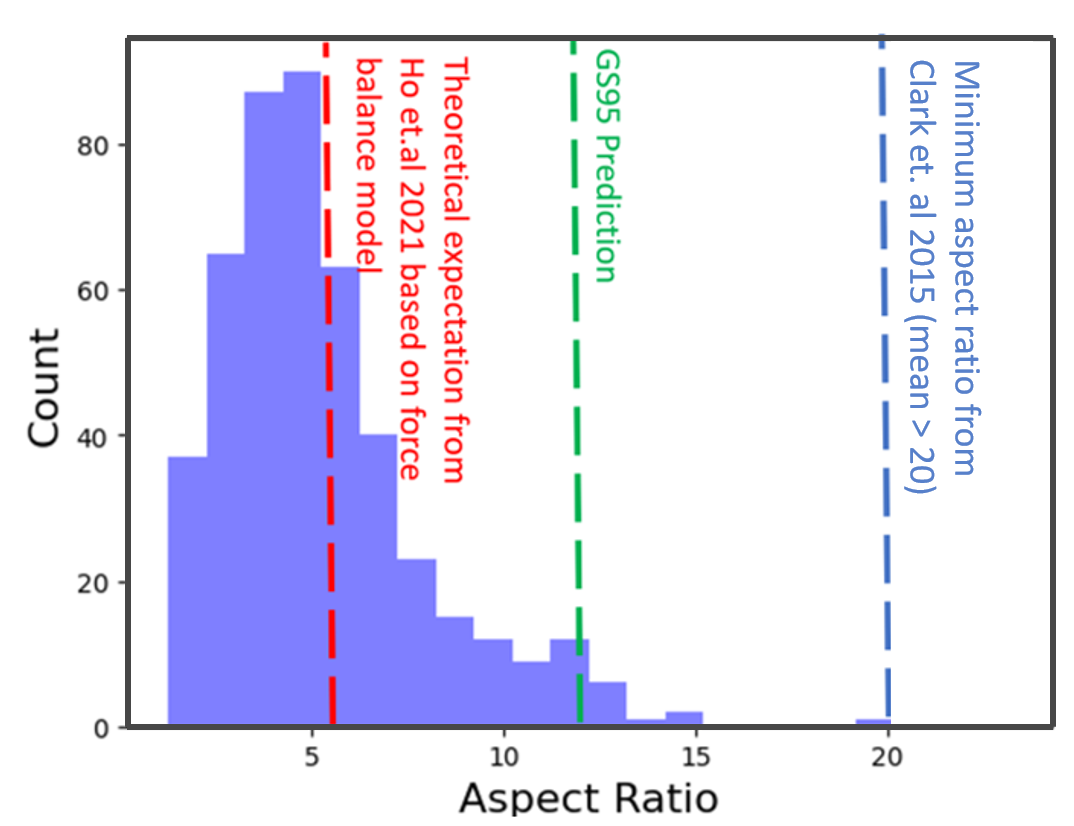}
    \caption{The aspect ratio distribution of {\avi filamentary structure from multi-phase simulation \citep{instability}, with the prediction from GS95 green line), the measured ratio by the RHT technique (blue line, \cite{2014ApJ...789...82C}), and the prediction including the effect from UNM (red line, \cite{instability}). It is therefore postulated that the elongated features in the sky are not real CNM and are not governed by GS95.}}
    \label{fig:fig3}
\end{figure}

\subsection{The main question of this review: Are filaments in the ISM actually physical?}

Before addressing the above questions, it would be fruitful to ask a simpler question: {\bf Are the observed filamentary features real or a computational artifacts?} Interpretation of these magnetically aligned, elongated over-dense features is widely discussed in the literature (See \citealt{2021A&A...654A..91K}), but understanding the nature of their physical properties, including their origin, stability, and fate, is crucial for comprehending the full physics of multiphase interstellar media, particularly in shaping and controlling the star formation rate in molecular clouds. Furthermore, how the more ambient HI gases at scales of kpc form molecular clouds at pcs, and subsequently stars in sub-pcs, remains an open question for the community.

Readers should note that while this review attempts to be as comprehensive as possible, the analysis of filaments has rapidly evolved since the renowned filament review by \cite{2014prpl.conf...27A}.  Machine learning vision algorithms are much more powerful in 2024 than they were in 2014. We will include as many important results as possible up to the authors' knowledge, with essential ideas covered and extended literature for readers to follow up.

\phantomsection 
\section{Filament identification and chatacteristic: An observational perspective}\label{sec2}

A key aspect of this review is to provide readers with the limitations of and caveats on characterising properties of filaments from observations (or more formally, a projected image from telescope observations) to infer the underlying physics, in which to compare with the output of simulations with a certain recipe that contains certain fundamental physics. To answer the central questions outlined in the previous sections demands a review of the state of the art in the identification methods of 'filamentary' structures in the heavens and the underlying inferred physics. Despite this review focusing on CNM filament, one of the tensions in comparing observations and simulation finds, independent of the interstellar phase of converters, is that the physical quantities reported from observational studies are structured 'post-processed' by certain methods that may already involve a certain underlying restriction on the meaning of 'filament' and the corresponding physics. However, simulations studied mainly concern reviewing the physics and do not normally take care of observational properties such as instrumentation artefacts, noise properties, and post-process methods, which could contaminate or even alter the resulting products, which led to an unfair comparison. In Section 2.1, we first review and classify different classes of filament identification approaches and what kind of properties these observed 'filaments' connect to the underlying physics. We then review important observational findings toward 'filaments' in the CNM. Here, we refer the reader to chapters from PPVI by Andre and Li and a chapter by Henshaw in PPVII for focused studies of the observational findings on filaments in the molecular clouds (i.e., molecular hydrogen-dominated phase). In particular, we review certain key observational metrics, such as the autocorrelation function, reduction factors (a.k.a alignment measures), model fitting to the radial density profile, and power spectrum, and their connection to the physics they result in such quantities.  While this review focuses on reviewing the fundamental regulating physics of structure formation and evolution in the multi-phase interstellar medium, a particular frontier in the aspect is the magnetic imprint on the gas kinematics and opens up a window to probe the interactions between two regulating mechanisms: magnetic field and turbulence. Filaments are in the spotlight as these commonly seen elongated structures or 'dark lanes' extending from the galactic plane are products of the interplay between the magnetic field and turbulence. Hence, the observed magnetic field properties in CNM provide important information. Confront and evaluate the reality of structures identified by different methods, but not without caveat. We see projected quantities from a 3-D universe, in which many underlying physics are complicated with the projection effects. Hence, it is worth reviewing and summarising the state of the art of projection and de-projection algorithms that allow us to have a glimpse of actual interactions. Here, we point the reader to Frontier's review by Thani on the state of probing 3-D magnetic field structures, as here we intend to focus on the method of converting between 2-D to 3-D, as observations are 2-D images. A spectral line cube is essentially a 2-D map, but also with the line of sight velocities information, which provides extra information for re-projection (see review by Tahani  et al. 2022). We end this section by stressing the essential need for synthetic observations and related visualisation as this allows a more realistic comparison between simulations that do not care about observational limitations but allows evaluation of the impact of observation parameters on the underlying observables or metrics that are the 'bridges' between theories, which is the main focus of next section the corresponding defining observables. Our goal is to articulate the current state in bridging observations and theories (via predictions based from simulation studies), and to address certain important aspects in order to make progresses in understanding the nature of 'filaments' in the CNM. This review is not intended to cover all aspects but to review the current status and identify gaps that require further studies in the next years. we recommend readers  to other reviews on similar topics but with different focuses to have a more comprehensive picture of the topic {e.g. PPVII chapters by \citealt{2023ASPC..534..153H} and \citealt{2023ASPC..534..233P}).

 \subsection{Methods in filament identification}
We approach the definition of "filaments" in an astrophysical setting by considering the following set of questions:
\begin{enumerate}
\item Are the visual features defined via computer identification algorithms (e.g., Unsharp mask (UNM), Rolling Hough Transform (RHT)) or based on some observational mechanisms (e.g., Velocity caustics)?
\item Are filaments fully continuous, partly discrete, or web-like? Do we expect sub-filaments within filaments?
\item Can projection effects misleadingly make 2D sheet-like structures appear as 1D filaments? (see, e.g., \citealt{MO07})
\end{enumerate}

Table 1 summarises most of the commonly used algorithms to identify filaments. In Figure 4, we also present some examples of 'filaments' defined by a few of the selected algorithms.

\begin{table}
\centering
\begin{tabular}{@{}ll@{}}
\toprule
Publication & Technique/Algorithm/Package \\ 
\hline
\cite{2008ApJ...679.1338R} & { Dendrogram}\\
\cite{2011MNRAS.414..350S} & { DisPerSe}\\
\cite{2013ApJ...774..128S} & { Intensity Gradient}\footnote{Readers should not confuse the method proposed by \citeauthor{2013ApJ...774..128S} and the "Velocity Gradient Technique" \cite{GL17,YL17a,YL17b,LY18a,LY18b} which is dedicated to trace magnetic field orientations.}\\
\cite{2014ApJ...789...82C,2015PhRvL.115x1302C} & Rolling Hough Transform \\
\cite{2015MNRAS.452.3435K} & FilFinder \\
\cite{2016ApJ...821..117K} &{Unsharp Masking }\\
\cite{2016AnA...593A..58J} & Template matching \\ 
\cite{GL17} & Velocity Gradient Technique \\
\cite{Soler2020} &   Hessian Matrix Technique \\
\cite{2020AnA...637A..18B} & T-ReX\\
\cite{VDA}         &   Velocity Decomposition Algorithm  \\
\cite{2022arXiv220500683A} & MaLeFiSenta\\
\cite{2023arXiv230806641X} & CASI-2D\\
\bottomrule
\end{tabular}
\caption{Common techniques used in the ISM and star formation community to extract and identify "filaments" in the last decade. See Fig.\ref{fig:illus_1} for some of the visualizations. }
\label{tab:def} 
\end{table}

\begin{figure}[th]
  \centering
  \includegraphics[width=0.88\textwidth]{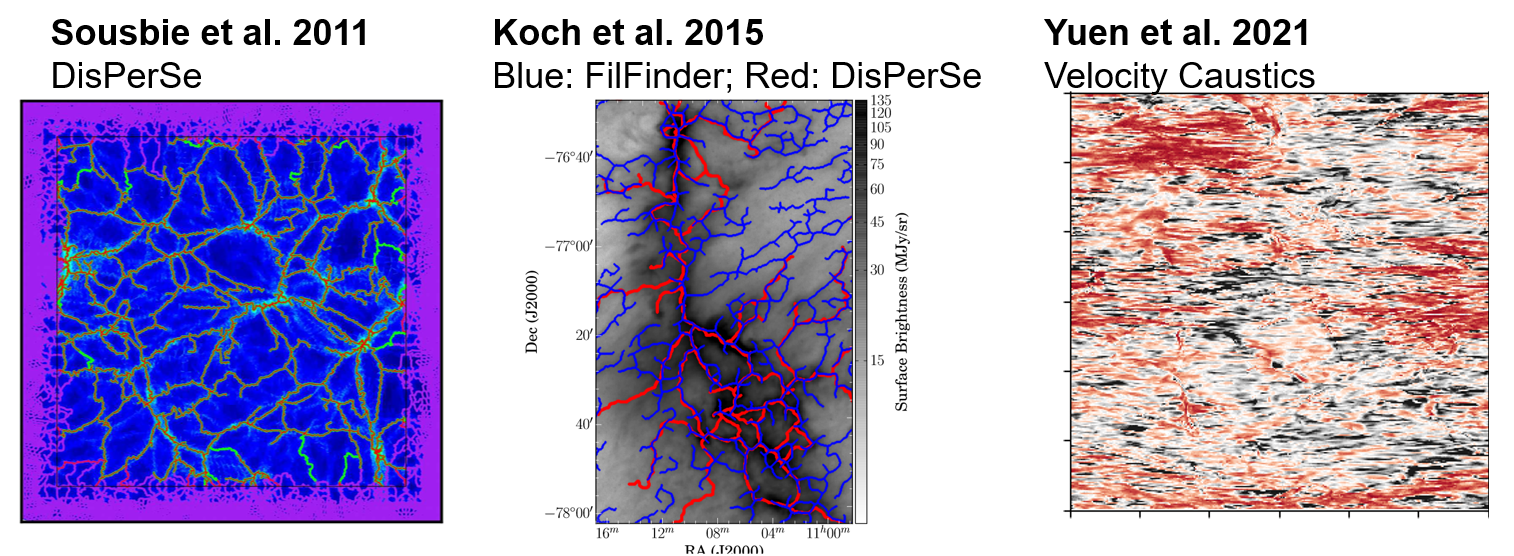}
  \caption{\label{fig:illus_1} Filaments identified from selected techniques (See Tab.\ref{tab:def}). From the left to right : DisPerSe method \citep{2011MNRAS.414..350S}, FilFilder \citep{2015MNRAS.452.3435K}  and Velocity Caustics Theory \citep{VDA}. Reproduction of the figures are permitted according to MNRAS guidelines.}
\end{figure}

These methods can be classified into two categories: (i) Image characterisations and (ii) machine learning techniques. Image characterisation techniques are subject to the impact on the intensity map from a given prescription of a real filamentary structure, which could be a dust continuum or an integrated intensity map of a specific molecular species. These prescriptions include continuous yet non-random over-dense regions with a net aspect ratio greater than unity (i.e. the eccentricity of the structure) and distinct start and endpoints. This term 'net' is important as it contains an underlying factor that there are reference regions that allow contrast to arise to single out these connected structures. For example, suppose the instrumentation noise is high enough to create a noise 'mask' that makes an integrated intensity map that a 2D uniform distribution could approximate. In that case, the filament will be convoluted with these noises and 'hide' within the 'mask' by losing out the contrast. However, actual observation data involve noises from the instrument and non-related structures. Thus, it is expected to set specific criteria to post confidence in the contract, which could also be known as the signal-to-noise (SNR) technique for filament identifications. These are the fundamental approaches for Dendrogram, Filfinder, and Unsharp Masking. Within this class of methods, a sub-class utilises more subjective approaches via statistical evaluations. For example, one can apply the intensity gradient and look for any continuous signal of contrast (intensity gradient reflects the region with the sharpest slope in the direction intensity profile along a given direction. Other approaches, such as the RHT and the Hessian matrix, are more sophisticated techniques to distill signals from the intensity map, \citealt{2019arXiv190403173Y}). Despite the wide variety of algorithms available, several common approaches can be summarized, with some illustrative examples listed in Table~\ref{tab:def}:

A clear caveat of image detection techniques is that it is highly subject to the prescription of a filament imposed on the intensity map: different prescriptions will yield different filaments. Such subjective prescription may be removed via Machine learning (e.g. CASI-2D). Machine learning techniques have emerged as powerful and time-efficient tools for filament identification by training on data of known filaments based on some criteria or models that implement certain physics (see next section). This can reduce bias compared to methods primarily relying on manual image analysis. However, machine learning models (e.g. \citealt{2022arXiv220500683A}, \citealt{2023arXiv230806641X})often require training datasets obtained from either observational data or simulations, which the quality of the input data can still limit.

The techniques mentioned above yield outputs that exhibit certain common properties. For instance, the orientation dispersion of filaments is strongly correlated with the Alfvenic Mach number \citep{dispersion,curvature}. Notably, diffuse, low-density filaments (considered low in density before the $HI$-$H_2$ transition) tend to align closely with the direction of the magnetic field \citep{2015PhRvL.115x1302C, GL17,YL17a}. In the following sections, we will summarize key findings related to the physical properties of HI and molecular cloud filaments. However, as we will discuss in Section 3, it is essential to have a correct theoretical foundation and a proper understanding of the physical nature of filaments when applying these algorithms.

\subsection{HI filaments and their relation to MHD turbulence}

Filamentary features are ubiquitous in both diffuse interstellar media and dense molecular clouds. Very early observations, both from absorption \citep{2001ApJ...551L.105H} and emission maps \citep{2006ApJ...652.1339M}, seem to indicate that the cold, dense ISM filaments are magnetically aligned. However, it was not until large-scale, high-resolution HI position-position-velocity (PPV) emission data (e.g., {\it GALFA}, \citealt{2018ApJS..234....2P}) and line-absorption data (e.g., 21SPONGE, \citealt{2018ApJS..238...14M}) became available that observers could gain a clear understanding of the turbulence and magnetic properties of cold, dense neutral hydrogen filaments.

The identification of filaments in HI data can usually be performed using three very popular techniques, in the order of their introduction: Rolling Hough Transform (RHT, 2014, \citealt{2014ApJ...789...82C}), which originated from the idea of particle track tracing in the 1950s and detects linear dense features in two-dimensional maps; Velocity Gradient Technique (VGT, \citealt{GL17,YL17a,LY18a}), which utilizes the first spatial derivative of velocity-related maps (see the first two equations of Eq. \ref{eq:velocity_map}) to identify features; and Hessian Matrix Method (2019, \citealt{2019A&A...622A.166S}), which evolved from the Intensity Gradient Method \citep{2013ApJ...774..128S} but attempts to perform second-order statistics. Despite employing different methods to identify filaments, these statistical models are all based on the same underlying principle: How {\bf magneto-hydrodynamic turbulence} interacts with thermodynamics.

Despite differences in methodology, a general trend among these methods follows: in the more diffuse interstellar media, the identified filaments are generally aligned with the underlying magnetic field directions \citep{2003ApJ...586.1067H,2015PhRvL.115x1302C,GL17,YL17a}. When denser clouds are measured, regardless of the phases, they tend to be perpendicular to the magnetic field direction \citep{2013ApJ...774..128S,YL17b}. Molecular cloud filaments will be discussed in the next section, and below, we shall delve into the turbulence aspect of the diffuse HI.

Diffuse interstellar media are known to be multiphase \citep{1977ApJ...218..148M,2017NJPh...19f5003K}, magnetized \citep{2001ApJ...551L.105H}, and turbulent \citep{2010ApJ...710..853C,spectrum}. These factors at large scales dominate the small-scale physics that is usually of more interest to the star-forming community, such as star formation \citep{2010A&A...512A..81F,2015Natur.520..518L}, ion-neutral damping \citep{2010ApJ...718..905L,2015ApJ...810...44X}, and outflow and stellar feedback \citep{2018ApJS..237...22S}. The diffuse nature of the ISM allows one to model the dynamics of gas and phases with simple physical models that contain essential information like turbulence with magnetic fields \citep{GA,curvature,2023ARA&A..61...19M}. 

The physics of interstellar filaments is highly connected to the properties of the "unstable neutral media" (UNM), which was previously considered an intermediate phase in the multiphase interstellar media \citep{1977ApJ...218..148M}. One of the most prominent features of filaments is their very high aspect ratio, frequently reported in observations. In Tab.~\ref{tab:filament_ar}, we have listed several examples in both diffuse HI and dense molecular clouds. It is evident that the long aspect ratio of filaments makes them unique quantifiers in observations.
\begin{table}
\centering
\begin{tabular}{@{}lll@{}}
\toprule
Publication & Aspect Ratio & Object \\ 
\hline
\multicolumn{3}{l}{\bf Diffuse ISM}\\
\cite{2022ApJ...941....6W} & $\sim$500 & H$\alpha$ from Synchrotron Emission \\
\cite{2014ApJ...789...82C} & $\sim$60 & Diffuse HI \\
\cite{2021ApJ...918L...2L} & 25 (5kpc), 5 (1kpc) & Cattail \\
\cite{2022AnA...657A...1S} & $\sim 30$ & Maggie \\
\multicolumn{3}{l}{\bf Molecular Cloud}\\
\cite{2021AnA...655L...2C} & $8 \sim 26$ & Molecular clouds from outer galaxy  \\ 
\cite{2020AnA...642A..87K} & $1.2 \pm 0.4$ & Hub-filament systems from Herschel survey \\ 
\cite{2019AnA...621A.130L} & $\sim 23$ & G351.776–0.52 \\
\cite{2020AnA...644A..27B} & $54$ & Musa \\
\cite{2010ApJ...719L.185J} & $150$ & Nessie Nebula \\
\cite{2020AnA...642A..76Z} & $4-8.4$ & California \\
\bottomrule
\end{tabular}
\caption{Filament aspect ratio from different observations. These selection of examples are non-exclusive.}
\label{tab:filament_ar}
\end{table}

It is well-known that certain types of turbulence exhibit universal relations in some statistical quantities. One of the most renowned is the Kolmogorov scaling \citep{K41}, which states that, for 3D isothermal/adiabatic, incompressible, hydrodynamic turbulence, the power spectrum follows a simple scaling law:
\begin{equation}
E(k) \propto k^{-5/3}
\end{equation}
Observationally, this scaling relation has been demonstrated from time to time. One of the most recent efforts is shown in Fig.~\ref{fig:spectrum_fig1}, where a very large power law is observed across almost six decades of length scales in the sky. { Recently, \cite{2023ApJ...949...30L} found results that seem to differ from that of Fig.~\ref{fig:spectrum_fig1}. However, \cite{2023ApJ...949...30L} focuses on regions that are gravitationally bound, and star-forming activities may impact the local turbulence energy spectrum. In fact, earlier studies \citep{2018MNRAS.477.4951L} have shown that the presence of strong gravitational collapse \citep{2015Natur.520..518L} will cause the turbulence spectrum to deviate from the standard \citeauthor{K41} cascade. The results presented in \citep{2018MNRAS.477.4951L} show that a universal spectral index is found across ISM from WNM to CNM.  }

\begin{figure}[th]
  \centering
  \includegraphics[width=0.88\textwidth]{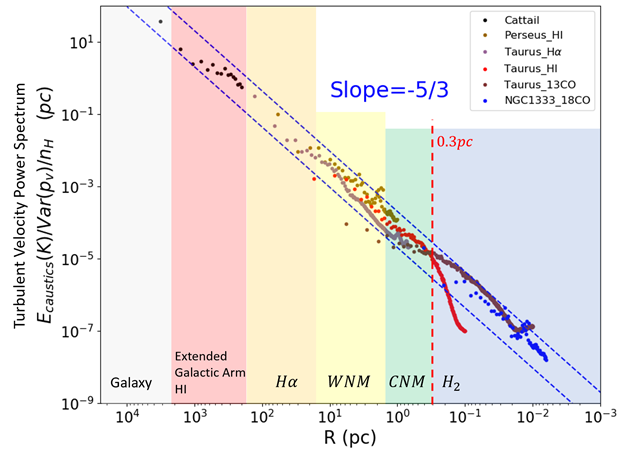}
  \caption{\label{fig:spectrum_fig1} The Big Power Kolmogorov Law for ISM velocity fluctuations spans from $10^4$ pc to $10^{-2}$ pc, as analyzed via \cite{VDA} using archival data of ISM observations (see \cite{spectrum} for detailed information). The universal cascade is evident, extending from turbulent velocities at the galactic arm scale down to star-forming hubs. To assist readers in recognizing the cascade, we have labelled the dominant scale for each phase and included auxiliary lines representing the Kolmogorov spectrum (the blue dashed lines). This information is derived from \cite{spectrum}.}
\end{figure}

Astrophysical turbulence is inherently magnetized and compressible, a topic that has been thoroughly investigated by numerous authors in the community (e.g., \citealt{CL03,2007ApJ...658..423K}). The presence of compressibility in turbulence adds complexity to our understanding of how the three fundamental MHD modes (Alfven, slow, and fast modes), each characterized by different spectra and anisotropies, behave in various astrophysical phenomena. For instance, it is widely believed that the Alfven and slow modes play a central role in aligning the cold neutral media with the magnetic field \citep{LY18a} and controlling the transport of heat and particles across magnetic fields \citep{2001ApJ...562L.129N,2021arXiv210801936M}. In contrast, fast modes are pivotal in the scattering and acceleration of cosmic rays (\citep{2002PhRvL..89B1102Y,2004ApJ...614..757Y,2005ThCFD..19..127C,2007MNRAS.378..245B}). The composition of these modes strongly depends on the driving parameters \citep{2020PhRvX..10c1021M}.

The magnetic component in multiphase turbulence provides additional directional guidance to the dynamics of ISM gases before reaching the H1-H2 conversion threshold \citep{2016A&A...595A..32B,2017ApJ...835..126B}. The morphology of ISM gases tends to be anisotropic along the local direction of the magnetic field, primarily due to the Goldreich-Sridhar cascade \citep{GS95}.

The fundamental idea behind Goldreich-Sridhar turbulence is as follows: When turbulence is strong \footnote{In this review, we do not delve deeply into the fundamental properties of magnetized turbulence. The classification of "weak" and "strong" turbulence is based on the relative amplitude of the linear and nonlinear frequencies (see, e.g., \citealt{Lazarian06,2022JPlPh..88e1501S}), assuming that non-local transport is small.}, it is postulated that the nonlinear time is of the same order of magnitude as the linear propagation time \citep{GS95}. In this scenario, the parallel and perpendicular length scales in the local frame of reference \citep{2000ApJ...539..273C} exhibit the following scale-dependent relationship:

\begin{equation}
l_\parallel \propto l_\perp^{2/3}
\end{equation}

A significant consequence of the Goldreich-Sridhar cascade is that, in cases where no additional heating and cooling processes are at play (\citealt{1983ApJ...270..511Z}), observables in interstellar media driven by MHD turbulence tend to exhibit a filamentary nature. This phenomenon was well observed in early simulations \citep{2002PhRvL..88x5001C,2010A&A...512A..81F} and is also evident in observations \citep{2008ApJ...680..420H}. As a result of this fundamental characteristic, it is widely believed that observational maps, whether linear (such as Column Density and Velocity Centroid, \citealt{2005ApJ...631..320E}) or nonlinear (like Velocity Channel Maps, also known as PPV slices, \citealt{LP00}), are inherently filamentary due to the filamentary nature of their 3D counterparts. Here are some of the more common observables:

\begin{equation}
\begin{aligned}
\textbf{Velocity Centroid: }& \quad C(\mathbf{X}) \propto \frac{\int_{\cal L} dz v (\mathbf{X},z) \rho (\mathbf{X},z)}{ \int_{\cal L} dz \rho (\mathbf{X},z)} \\
\textbf{Column Density: }& \quad I(\mathbf{X}) \propto \int_{\cal L} dz \rho (\mathbf{X},z)\\
\textbf{Velocity Channel: }& \quad p(\mathbf{X},v_0) \propto \int_{v\in v_0\pm \Delta v/2} dz \rho (\mathbf{X},z) \phi(v)
\end{aligned}
\label{eq:velocity_map}
\end{equation}

Here, $\phi(v)$ is the kinetic kernel that incorporates contributions from turbulence, shear, and thermal broadening \citep{VDA}. The prevalence of magnetized turbulence in the diffuse ISM allows the methods listed above to function both as identification techniques and as magnetic probes. We will delve into more technical aspects of how filamentary features in 3D multiphase interstellar media are translated into 2D representations in later sections.
Due to observational limitations, 2D observational maps (as described in Eq.~\ref{eq:velocity_map}) that appear filamentary are not always indicative of 3D filaments, despite the strong correlation observed in both observations \citep{2015AnA...584A..67B} and simulations \citep{2015ApJ...810..126C}. For example, strong shocks are readily visible in compressible magnetized turbulence, and when viewed from the side, they can appear filamentary. Various projection effects have been well-documented; you can refer to \cite{2013ApJ...777..173B} for more information. Moreover, when a 3D filament is projected using nontrivial observables like velocity channels, additional effects due to the velocity mapping \citep{LP00,2013ApJ...770..141B,2013ApJ...777..173B,VDA} come into play.

This additional effect can fragment real 3D density filaments that possess a line-of-sight velocity dispersion (typically nonzero for molecular clouds \citep{1981MNRAS.194..809L}) into pieces distributed across different channel maps. It can also combine two spatially separate 3D filaments with similar velocities into a single artificial filament in one channel map. We will discuss this nontrivial effect and its implications for several relevant observational findings in later sections.

It is crucial to emphasize that in the context of the diffuse ISM, filaments do not necessarily exhibit high density or intensity. In cases where gravity does not significantly concentrate high-density features, and when the plasma parameters are generally subsonic and sub-Alfvenic in the warm phase, the direct effect of the anisotropic MHD cascade primarily influences velocity \citep{Lazarian06}.

Density fluctuations are influenced by velocity fluctuations and can be viewed as a mirrored response to the anisotropic velocity fluctuations, both for regions of high and low density \citep{2007ApJ...658..423K}. However, as we mentioned in previous sections, MHD turbulence alone cannot fully explain the statistical features of ISM filaments observed in the sky \citep{2021A&A...654A..91K,instability}. We will delve into this further in Section 3. In fact, the properties of HI filaments are closely connected to the underlying phases of the surrounding region.

Observational studies have attempted to measure the statistics between the three phases by comparing 21 cm line absorption and emission profiles.\cite{2001ApJ...551L.105H} discovered that approximately 50\% of the multiphase gas falls within a temperature range of 500 to 5000 K, indicating the presence of significant amounts of thermally unstable phases later referred to as Unstable Neutral Media (UNM) within the observed ISM. More recent absorption line studies by \cite{2018ApJS..238...14M} suggest that 40\% of the gases are in the UNM. Specifically, at least 20\% of the total HI mass falls within the temperature range of 250 K to 1000 K, corresponding to the lower end of the UNM.  It's worth noting that absorption line studies are limited to detecting the lower temperature end of the HI.

As evident from \cite{2018ApJS..238...14M}, a significant portion of UNM falls into the self-absorption regime. Consequently, previous observational studies that primarily focused on the structures of HI (e.g., \citealt{2015PhRvL.115x1302C,2019ApJ...887..136C}) likely underestimated the mass fraction of UNM and CNM \citep{2022MNRAS.512.4765S}. Nonetheless, the above observations underscore the significance of UNM in the multiphase ISM, which shapes the morphology and statistical characteristics of filamentary features within it.

While we won't delve into the majority of the observed properties of molecular cloud filaments here, it's important to note that these filaments inherit their properties from their diffuse HI counterparts and still adhere to the same turbulence cascade \citep{1981MNRAS.194..809L,spectrum} until they reach scales that are small enough to enter the gravitational instability regime, which typically flattens the spectrum \citep{2018MNRAS.477.4951L}, or until they are influenced by "micro-physics." The intricate interplay between turbulence, gravity, magnetic fields, and micro-physical processes has given rise to extensive research linking star formation \citep{2017NatAs...1E.158L,2019ApJ...883..122L} to the turbulence properties of molecular clouds \citep{2018MNRAS.480.3916M}.

Surprisingly, molecular cloud filaments exhibit a wider range of aspect ratios than their diffuse counterparts. The lower half of Tab.\ref{tab:filament_ar} displays recent observations of filament aspect ratios. We caution readers that some of the "filaments" identified by observers are actually visually distinct and dynamically separated features that happen to result from the fragmentation of a single cloud. Therefore, the numbers reported in the literature should be interpreted with some caution. Fortunately, being isothermal and subject to strong turbulence allows for the quantification of aspect ratios based on simple plasma parameters, which we will discuss in Section 3.1. Notably, a study by \cite{2019ApJ...878..157X} attempts to use the Goldreich-Sridhar scaling to explain the morphology of isothermal filaments. It is relatively straightforward to assess whether the numbers in Tab.\ref{tab:filament_ar} are consistent with the Goldreich-Sridhar estimate by approximating the Alfvenic Mach number. In general, filaments in molecular clouds have following characteristics , but it is not the goal of this review to have an in-depth discussion (see, e.g., \citealt{2014prpl.conf...27A,2016ApJ...824..134F} and references therein, or Table \ref{tab:def}):

\begin{enumerate}
\item Typically, filaments are denser, confined by self-gravity, and isothermal features \citep{2014prpl.conf...27A}.
\item Filaments can undergo fragmentation due to various factors, such as Jeans instability or other physical processes that disrupt their stability \citep{2014ApJ...791..124G,2019MNRAS.490.3061V}.
\item The width of filaments range between $\sim 0.03~pc$ to approximately $\sim0.15$ pc \citep{2016ApJ...824..134F}. However, it is still a debatable question whether there is an universal filament width across different scales and environments. 
\item A "Plummer-like" density profile $\rho \propto (1+r^2)^{-1}$ is often observed \citep{2019AnA...621A..42A}.
\item Various processes can create filamentary-like features, including outflows, stellar feedback, shocks from bubbles, or other micro-turbulence (e.g.  \citealt{2023ASPC..534..233P,2023ASPC..534..153H}).
\end{enumerate}

\subsection{Filamentary structures seen in other context }

While the term "filaments" is primarily used in the context of the diffuse interstellar medium and the star-forming community, the same concept has been extended to other types of measurements. Filamentary-like features have been observed in early measurements of synchrotron intensity and polarization maps \citep{2011Natur.478..214G} and even in cosmological research and intra-cluster medium (ICM). Here, we briefly review the literature that discusses filaments with scales smaller than the typical turbulence injection scale \citep{2010ApJ...710..853C}. Filaments originating on larger scales typically arise from other physical processes, such as anisotropic thermal conduction in the case of the ICM \citep{2012ApJ...749...45C}. Some examples include \citep{2012ApJ...754..122K,2020ApJ...889L...1L}.

Earlier observations of nonthermal emissions (e.g., \citealt{1984Natur.310..557Y}) revealed extensive filament-like features associated with poloidal magnetic flux tubes extending from the Galactic center. More recently, \cite{2011Natur.478..214G} observed a significant network of filaments in synchrotron polarization maps, which were later identified in numerical simulations as indicators of the sonic Mach number \citep{2013ApJ...771..123B} and as probes of the magnetic field \citep{LYLC,LY18b}. In fact, when analyzing observational maps of the diffuse interstellar medium using gradient-based algorithms, it is quite common to observe a network of filament-like features. These features can be easily reproduced by even the simplest high-resolution magnetohydrodynamic simulations, regardless of the Mach number \citep{GA}. However, other mechanisms can create filamentary features in synchrotron polarization maps that are larger than the turbulence correlation length or even the injection scale typical of the interstellar medium \citep{2010ApJ...710..853C}. For instance, the "North Polar Spur" is believed to have formed due to the influence of a local bubble \citep{2021ApJ...923...58W}.

There are other notable methods for detecting filaments, including dust emission and Faraday rotation maps. Dust polarization arises from the emission of non-spherical grains that align with their long axes perpendicular to the ambient magnetic field (see \citealt{2015ARA&A..53..501A}). Similarly, polarization of starlight results from the differential extinction of aligned grains. Dust alignment is generally believed to occur due to radiative torques (RATs) (see \citealt{1976Ap&SS..43..257D,1996AAS...189.1602D}). The theory of RAT alignment is based on the analytical model presented in \cite{2007MNRAS.378..910L} and further studies, such as \cite{2008MNRAS.388..117H,2016ApJ...831..159H}. Dust emission maps in the diffuse to moderately dense parts of molecular clouds are roughly linearly proportional to the actual gas density and can be treated as an alternative density tracer in observations.

Faraday rotation maps share very similar characteristics to synchrotron emission maps. In general, synchrotron emission depends on the distribution of relativistic electrons:
\begin{equation}
N_e({\cal E})d{\cal E} \sim {\cal E}^{\alpha} d{\cal E},
\end{equation}
with the intensity of the synchrotron emission given by:
\begin{equation}
I_{\text{sync}}({\bf X}) \propto \int dz B_{\perp}^\gamma({\bf x}),
\end{equation}
where ${\bf X} = (x,y)$ is the 2D position in the sky (POS) vector and $B_{\perp} = \sqrt{B_x^2 + B_y^2}$ represents the magnitude of the magnetic field perpendicular to the line of sight (LOS) along the $z$-direction. In general, $\gamma = (\alpha+1)/2$ is a fractional power, and LP12 demonstrated that the statistics of $I(\alpha)$ are similar to those of $I(\alpha=3)$. Therefore, it suffices to discuss the statistical properties in the case of $\alpha=3$.

According to \cite{LP12}, the synchrotron complex polarization function with Faraday rotation can be expressed as follows:
\begin{equation}
P_{\text{synch}}({\bf R}) = \int dz \epsilon_{\text{synch}} \rho_{\text{rel}} B^2 e^{2i\left(\theta({\bf R},z) + C\lambda^2\Phi(R,z)\right)}
\end{equation}
Here, $\epsilon_{\text{synch}}$ represents the emissivity of synchrotron radiation, and $\Phi(R,z)$ is the Faraday Rotation Measure (RM):
\begin{equation}
\Phi(R,z) = \int_\infty^z dz' \frac{1}{\sqrt{4\pi}} \rho_{\text{thermal}}({\bf R},z) B_z({\bf R},z) \text{ rad m}^{-2}
\end{equation}
In this equation, $\rho_{\text{rel}}$ is the relativistic electron density, $\rho_{\text{thermal}}$ is the thermal electron density, and $B_z$ is the magnetic field component along the line of sight (LOS). The C-factor is approximately 0.81 \citep{LYLC}.

For frequencies lower than around 1 GHz, the Faraday Rotation Measure can exceed $2\pi$, which results in a phase shift that causes a complete loss of information from the source. This leads to the concept of Faraday screening, which determines the maximum line-of-sight distance that can be observed in synchrotron emissions in the presence of a line-of-sight magnetic field. In the case of sub-Alfvénic turbulence, the source term $P_i = \rho_{\text{rel}}\exp(2i\theta({\bf R},z))$ is dominated by the mean magnetic field rather than the fluctuating field.

Two regimes, strong and weak Faraday Rotation, depend on whether the ratio of the scale sampled by polarization to the size of the emitting region, denoted as $L_{\text{eff}}/L$, is smaller (strong) or larger (weak) than unity:
\begin{equation}
\frac{L_{\text{eff}}}{L} \sim \frac{1}{\lambda^2L} \frac{1}{\phi}
\end{equation}
In this equation, $\phi$ is defined as $max(\sqrt{2} \sigma_\phi, \bar{\Phi})$, where $\sigma_\phi$ represents the dispersion of the random magnetic field and $\bar{\Phi}$ is the average Faraday Rotation Measure. Given that the Faraday screen is typically non-zero in modern observations (e.g., LoFar), interpreting filamentary features in low-frequency maps requires exceptional care.

\section{Theory of the formation of filamentary features in multi-phase interstellar media}
\label{sec3}
In Section 1, we listed two fundamental questions related to cold neutral media filaments. From the above, we know how filaments are measured. However, it is still not clear why filaments in the ISM have different properties than those in simulations, nor the nature of these filaments. To understand why and how filaments are formed under the complex interplay of multiphase, multi-physics phenomena, we will need to delve into the theories and find insights into the effects of macro- (i.e., universal processes that affect all scales, e.g., turbulence, multiphase heating and cooling, etc.) and micro-physics (e.g., gravity on small scales, chemical evolution, HI shielding, etc.). These different physical processes act together to form the observed multiphase interstellar media. Understanding how physics controls the gas and phases is, therefore, the first step in answering the questions from Section 1.

The main goal of this review is to discuss the nature and appearance of filaments seen in the interstellar medium. Dense filaments in molecular clouds where the natural width ($\sim 0.1$ pc, \citealt{2014prpl.conf...27A,2016ApJ...824..134F}), orientation (preferentially perpendicular to the B-field when density is high, \citep{2013ApJ...774..128S,2013MNRAS.436.3707L, 2019ApJ...878..157X}), and network-like structure has been well studied. However, their diffuse ISM counterpart is much less well understood.

Various models have been proposed to describe the formation of the diffuse filament:
\begin{enumerate}
\item Turbulence, which due to compression \citep{2007ApJ...661..972P,2010A&A...512A..81F}, stretching, and mixing in the presence of a magnetic field \citep{CL03,2007ApJ...658..423K} naturally produce filamentary features.
\item Other non-MHD physical mechanisms, e.g., heating and cooling \citep{2013A&A...556A.153H,2017NJPh...19f5003K,instability}, "turbulent conduction and mixing" \citep{2004JKAS...37..557C}, chemical evolution \citep{2023arXiv230504965G}, and cosmic ray support \citep{2019A&A...622A.143C}, which can significantly reshape the filamentary features, particularly due to the ubiquity of unstable neutral media (UNM, \citealt{2018ApJS..238...14M}).
\item Gravity, which plays a small role in confining the geometry of filaments \citep{2018MNRAS.476.4932V}.
\end{enumerate}
As a result of these filaments formation mechanicsm from different ISM phases can exhibit very different visual properties, which we will discuss in \S 3.1 and \S 3.2 separately

The study of multiphase interstellar filaments was highly limited to small-scale simulations and hand-waving arguments until very recently when it became numerically feasible to perform large-scale simulations that could realistically model phase transitions and resolve complex physics. Unlike isothermal simulations, the use of direct numerical simulations (DNS) on multiphase media only began in the last two decades. The thermal instability relation to turbulence and the analytic expression of the realistic cooling function have only been explored recently through the efforts of small-scale simulations \citep{2002ApJ...564L..97K,2002ApJ...569L.127K,2004ApJ...602L..25K}. The modern understanding of multiphase media thermodynamics includes a density-independent heating term and a temperature-dependent cooling term due to the Ly-$\alpha$ and CII lines, respectively. With the introduction of turbulence and a realistic cooling function, numerical studies have observed that a fraction of intermediate unstable neutral media (UNM) could exist stably throughout the evolution of hydrodynamical simulations \citep{2002ApJ...569L.127K,2014ApJ...786...64K}. This finding is further supported by subsequent MHD simulations with higher resolutions \citep{2017NJPh...19f5003K}. This UNM could significantly influence the dynamics of diffuse interstellar filaments \citep{instability}, which are already influenced by the ubiquitous magnetized turbulence \citep{GS95}. In this chapter, we explore why filaments can maintain a large aspect ratio, supported by numerical evidence from the community, and why perpendicular filaments in the ISM are not seen in observations.

\subsection{Goldreich-Sridhar Turbulence}

GS95 scaling is one of the most important scalings in modern MHD turbulence theory. It was originally developed to explain why it is not possible to have a three-wave interaction in a turbulent cascade. The classical paper by \cite{GS95} later became the foundation for the numerical exploration of compressible MHD turbulence.

It is important to note that the GS95 theory was originally developed for incompressible turbulence with an Alfvenic Mach number $M_A(=V_{inj}/V_A)=1$. In the case of sub-Alfvenic turbulence ($M_A<1$), there is an range of length scales where strong turbulence is not yet in play, meaning that the nonlinear time is still longer than the linear time. The turbulence in the range from the injection scale $L_{inj}$ to the transition scale
\begin{equation}
l_{trans}=L_{inj}M_A^2
\label{trans}
\end{equation}
is termed weak Alfvenic turbulence \footnote{It is worth emphasizing that, including Eq.\ref{la}, the transition scale assumes that the largest energy is at the injection scale. In reality, every Fourier mode has its own transition scale, but since the injection scale has the most energy, numerical simulations might appear to have a transition regime at Eq.\ref{trans} or \ref{la}. However, the actual transition should be a range of scales.}. This type of turbulence keeps the $l_{\parallel}$ scale constant while the velocities change as $v_\perp\approx V_L (l_\perp/L_{inj})^{1/2}$ \citep{LV99}. The cascading results in a change in the perpendicular scale of eddies, $l_\perp$, only. With the decrease of $l_\perp$, the turbulent velocities, $v_\perp$, decrease. Nevertheless, the strength of non-linear interactions of Alfvenic wave packets increases (see \citealt{Lazarian16}). Eventually, at the scale $l_{trans}$, the turbulence transitions into the strong regime, which obeys the critical balance condition proposed by GS95.

The situations where $l_{trans}$ is less than the turbulence dissipation scale $l_{diss}$ would require an unrealistically small Alfvenic Mach number $M_A$ for typical ISM conditions. Therefore, in most cases, ISM turbulence transitions to the strong regime. If the telescope resolution is sufficient to resolve scales less than $l_{trans}$, we should observe the signature of strong turbulence in observations. For injection of roughly 100 pc\citep{2009ApJ...693.1074C} and an Alfvenic Mach number of $\sim 0.5$ \citep{Kalberla2015}, the corresponding length scale is $\sim 25pc$. Readers, however, should be reminded that the actual Mach number in ISM changes significantly due to phase transitions, and therefore the transition is believed to be a broadband of length scales.

The exact scaling was developed and studied extensively later by \citeauthor{LV99}(\citeyear{LV99}, see also \citealt{2000ApJ...539..273C,2001ApJ...554.1175M}), among others:
\begin{equation}
\begin{aligned}
\text{\bf Critical Balance } &: v_\perp/l_\perp &= v_A/l_\parallel\\
\text{\bf Constant Energy Cascade} \perp B &: \frac{v_\perp^3}{l_\perp} &=\text{const}
\end{aligned}
\end{equation}
Here, $v_{\parallel,\perp}$ denotes the velocity fluctuations parallel and perpendicular to the magnetic field, while $l_{\parallel,\perp}$ denotes the length scales parallel and perpendicular to the magnetic field. The first equation postulates the equivalence of nonlinear and linear timescales (See, e.g. \citealt{2015MNRAS.449L..77M,2022JPlPh..88e1501S,4DFFT_p1} for more recent discussions and experiments), while the second condition arises because MHD turbulence appears hydrodynamic when we consider only perpendicular directional motion\footnote{This is only possible when we consider that flux tubes are mobile under some sort of mechanism, e.g., \cite{LV99} enables the reconnection diffusion mechanism to be in play}. By combining these two conditions, we can derive the length scales discussed earlier:
\begin{equation}
l_\parallel = L_{inj}^{1/3}M_A^{-4/3}l_\perp^{2/3}
\label{eq:gs95_full}
\end{equation}
For velocity $v_l$ we use the critical balance relation ($\omega_{nl}\sim v_\perp /l_\perp \sim \omega_A = v_A/l_\parallel$):
\begin{equation}
\begin{aligned}
v_l &= v_A \frac{l_\perp}{l_\parallel} \\
&= v_A M_A^{4/3} \left(\frac{l_\perp}{L_{inj}} \right)^{1/3} \\
&= v_{inj} M_A^{1/3} \left(\frac{l_\perp}{L_{inj}} \right)^{1/3}
\end{aligned}
\label{eq:gs95_v}
\end{equation}
(see \citealt{2015ApJ...810...44X}).\footnote{Readers should be cautious in their choice of symbols. By "amplitudes," we actually mean the dispersion of velocity in real space since the Goldreich-Sridhar type argument was performed in Fourier space, and the inverse Fourier transform of squared Fourier amplitudes yields correlation functions.}. These scalings are crucial for interpreting MHD turbulence in observations since GS95 scaling suggests a natural anisotropy parallel to the direction of the magnetic field. Readers should be careful to note that the filaments referred to here are velocity or magnetic field fluctuations in incompressible turbulence, as density remains approximately constant over the course of dynamics. Treating density as an exact constant definitely does not apply to the multiphase media. In the next section we discuss how these scaling relations are modified in compressible turbulence.

The velocity and magnetic field striations created by strong Alfvenic turbulence provide guidance on how the morphology in 3D or 2D maps is correlated with the intrinsic statistics of MHD turbulence. Suppose we make an observation of "high velocity intensity" in a 3D cube by a simple intensity threshold at $v_0$. Then, combining Eqs. \ref{eq:gs95_full} and \ref{eq:gs95_v}, we can derive an expectation for the anisotropy of filaments defined by this threshold:
\begin{equation}
\text{Aspect ratio} = \frac{l_\parallel}{l_\perp}({v_l=v_0}) = \frac{v_{inj}}{v_0 M_A} = M_A^{-4/3}\left(\frac{L_{inj}}{l_\perp}\right)^{1/3}
\label{eq:ar_gs95}
\end{equation}
On the other hand, if one defines filaments via edge detection (e.g., Unsharp Mask, Rolling Hough Transform) with a definite $\Delta x$, then the aspect ratio is actually the derivative of Eq. \ref{eq:ar_gs95} with respect to $l_\perp$ with a correction factor that depends on the ratio $l_\perp/\Delta x$, as the edges of filaments are considered as two filaments in edge detection algorithms\citep{2019arXiv190403173Y}. The correction factor arises from the fact that a thick filament can be misclassified as two very thin filaments. While the length of the filament is not altered via Eq. \ref{eq:gs95_full}, the width is defined by the user-dependent choice $\Delta x$. In this case, one is not measuring the aspect ratio of the filament but rather a mixture of aspect ratios of thin filaments and the edges of thick filaments. Statistically, this increases the aspect ratio of the processed maps \citep{2019arXiv190403173Y}. We can quantitatively describe the situation as follows:
\begin{equation}
\text{Aspect ratio} = \frac{1}{3}M_A^{-4/3}\left(\frac{L_{inj}}{l_\perp}\right)^{4/3}\times
\begin{cases}
1 \quad \quad & (2\Delta x>l_\perp)\\
l_\perp/2\Delta x \quad & (l_\perp > 2\Delta x)
\end{cases}
\label{eq:ar_gs952}
\end{equation}
In other words, the "aspect ratio" of filaments, even in the simplest case, has obvious dependencies on how it is defined.

The projection of filaments into 2D space is much more complex. In Section 3.3 we discuss the channel map effect, which was first systematically studied by \cite{LP00}. However, here we first discuss the global-to-local frame transformation effect, first proposed by \cite{2000ApJ...539..273C} (See also \citealt{2000PhRvL..85.4656C,CL03,leakage,4DFFT_p1}). All the formulae above describe a filament in the local frame of reference. However, when filaments are projected onto a 2D plane, the frame effect has to be taken into account. In the following, we use the same mindset as \cite{leakage} in discussing how the filaments are projected onto the 2D plane, assuming we are computing the unnormalized velocity centroid ($C \sim \int dz v(x,y,z)$). For simplicity, we will also assume the mean magnetic field is perpendicular to the line of sight. We have to remind the reader that the anisotropy of the projection of a particular component of MHD variable depends on contributions from both the scaling relation (Eq. \ref{eq:gs95_full}) and the "tensor anisotropy" \citep{leakage}. For the tensor anisotropy, it is out of scope of the current review.

In the global frame of reference, the scale-dependent anisotropy is lost (see, e.g., \citealt{2000ApJ...539..273C,2007ApJ...658..423K}). This is because for filaments with dimensions $L\times H$ where $L\gg H$, its magnetic field wanders relative to the mean field with an angle dispersion of $\delta \theta \propto M_A$ \citep{leakage}. The global frame anisotropy is therefore determined by the most probable length and width measured with respect to the global magnetic field direction, which is independent of the filament dimension in the local frame of reference. The most probable length for a statistically sufficient collection of wandering filaments is $L$, while the probable width is $L\sin\theta$, in small angle:
\begin{equation}
\text{Aspect ratio} \propto M_A^{-1}
\label{eq:aspect_gf}
\end{equation}
which allows one to observe a different dependence on $M_A$ compared to Eq. \ref{eq:ar_gs952}. However, the length of the observed filament is still relevant to the intrinsic turbulent properties given by Eq. \ref{eq:gs95_full}. To measure the true dimensions of filaments, one requires the collection of filaments along the line of sight to be statistically sufficient, and as a result, individual imprints of 3D filaments can be separable via vision identification algorithms, and we revert to Eq. \ref{eq:ar_gs95} or Eq. \ref{eq:ar_gs952} depending on the type of algorithms one uses 

Readers should be reminded that we are discussing velocity or magnetic field striations in the incompressible limits. Fluctuations of velocity and magnetic fields are symmetric about the mean value as long as there are no other physical processes favoring the collection of either high or low intensity values. As a result, the anisotropy of low-intensity pixels induced by velocity or magnetic field fluctuations also carries statistical meaning in the GS95 framework, which is exactly the case for velocity channel or synchrotron polarization maps.

It should also be noted that the presence of strong turbulence allows magnetic reconnection via stochastic diffusion to happen \citep{LV99}. A recent proposal \cite{2020AnA...642A..87K} suggests that the formation of Tiny Scale Atomic Structure (TSAS, \citealt{2003ApJ...586.1067H}) could originate from reconnection. Some evidence for this is reported in numerical simulations \citep{2023ApJ...949L...5F}, even though there they refer to plasmoid instability in terms of the reconnection mechanism. We elect to discuss the possibility of reconnection-driven processes in a later publication and invite readers to read the review \citep{2020PhPl...27a2305L}.

\subsubsection{Compressible MHD turbulence}
\label{chap1.comp}

The first correction of the model above (c.f. Eqs. \ref{eq:ar_gs95}, \ref{eq:ar_gs952}, \ref{eq:aspect_gf}) for filaments in Table \ref{tab:filament_ar} is to include compressibility. Here we briefly summarize the scaling laws {\it in the local frame of reference} for compressible MHD turbulence as done in \cite{LY18a}. If the energy is injected { with the injection velocity $V_L$ that is} less than the Alfven speed { $V_A$}, the turbulence is {\it sub-Alfvenic}. In the opposite case, it is {\it super-Alfvenic}.Turbulence scaling for different regimes can be found in Table \ref{tab:regimes}. We now briefly describe the regimes in Tab. \ref{tab:regimes} and how GS95 sclaings are modified in compressible flows. A more extensive discussion can be found in the review by \cite{2013SSRv..178..163B}.

\begin{table}[t]
\small
\caption{Regimes and ranges of MHD turbulence. \label{tab:regimes} From \cite{ch5}}
\centering
\begin{tabular}{ccccc}
\hline
\hline
Type                        & Injection                                                 &  Range   & Motion & Ways\\
of MHD turbulence  & velocity                                                   & of scales & type         & of study\\
\hline
Weak                       & $V_L<V_A$ & $[L_{inj}, l_{trans}]$          & wave-like & analytical\\
\hline
Strong                      &                      &                                        &                 &                \\
sub-Alfv\'{e}nic            &  $V_L<V_A$ & $[l_{trans}, l_{diss}]$ & eddy-like & numerical \\
\hline
Strong                    &                        &                                          &                 &                   \\
super-Alfv\'{e}nic       & $V_L > V_A$ & $[l_A, l_{diss}]$                    & eddy-like & numerical \\
\hline
& & & \\
\multicolumn{5}{l}{\footnotesize{$L_{inj}$ and $l_{diss}$ are injection and dissipation scales, respectively}}\\
\multicolumn{5}{l}{\footnotesize{$M_A\equiv v_{inj}/v_A$, $l_{trans}=L_{inj}M_A^2$ for $M_A<1$ and $l_A=L_{inj}M_A^{-3}$ for $M_A>1$.} }\\
\end{tabular}
\end{table}

In the case of compressible turbulence, the MHD modes can be separated into {\it to at least}\footnote{Different from most of the ISM literature, we emphasize the presence of zero temporal frequency entropy modes since the ISM temperature profile is not simple (See Fig. \ref{fig:fig2}). } three MHD modes: Alfven, slow, and fast modes, where only the last two modes contribute to density fluctuations. These magnetosonic modes behave very differently than of Alfven modes. 

Let us first discuss slow modes. In the case of high plasma $\beta = 2M_A^2/M_s^2$ subsonic turbulence, density fluctuations are enslaved by their velocity counterpart \citep{Lazarian06}, and thus the aspect ratio of density fluctuations can be modeled by the argument made for velocity striations in Eq. \ref{eq:gs95_full}. The properties of low $\beta$ slow modes are much more complicated; earlier studies suggested that the anisotropy imprint of low $\beta$ slow mode is actually weaker than that of the Alfvenic counterpart \citep{CL03, 2007ApJ...658..423K, 2010ApJ...720..742K}. The most extensive recent analysis is \cite{2020PhRvX..10c1021M}, where Fig. 9 suggests that the low $\beta$ anisotropy is somewhere between GS95 and isotropic and systematically weaker than that of the Alfvenic mode. Nonetheless the anisotropy of slow modes is still believed to have a strong correlation to GS95 in small scales, and the same toolset that we discussed in Eq. \ref{eq:ar_gs95} can be used for slow modes {\it as long as the anisotropy is scale-dependent}.

Fast modes are much more complicated and require extensive studies since they are highly related to the density structures formed in the interstellar media (See also statistical analysis from \cite{2020PhRvX..10c1021M}, and recent satellite observations from \cite{2023arXiv230106709Z, 2023arXiv230512507Z}). In subsonic regimes, fast modes are isotropic and exhibit a $-3/2$ or $-2$ spectrum \citep{2020PhRvX..10c1021M}. Fast mode in high $\beta$ cases is the main contributor of density fluctuations, but it is usually not important unless the sonic Mach number is over unity. Shock-induced filaments can be arbitrarily compressed in the case of isothermal media from classical analysis, but can only be compressed by a factor of $(n+1)/(n-1)$ for adiabatic cases where $n$ is the adiabatic index (See a comprehensive analysis from \cite{2019ApJ...878..157X}\footnote{We have to specifically note that when $M_s,M_A>1$, there is a maximum compression ratio that depends on the Mach numbers.}). We can see that the thermal properties of turbulent plasma, characterized by the adiabatic index \( n \), are crucial in explaining the aspect ratio of shock-driven filaments. In the context of a multi-phase interstellar medium (ISM), the adiabatic index could be $n < 0$ for unstable natural media, due to the strong cooling effect. This scenario promotes a pressure gradient directed toward the dense center, thereby favoring the formation of shock-induced turbulence, even when the sonic Mach number \( M_s \) is only slightly greater than unity.

Fast mode in supersonic low $\beta$ is the most complicated case of all kinds. Interstellar media is almost always low $\beta$ when the length scale is above the ion-neutral decoupling scale ($\sim$ 0.1 pc), \citealt{2015ApJ...810...44X}). However, not until the ISM enters the lower end of the unstable phase or the cold phase will the sonic Mach number become larger than 1. In the case of molecular clouds, the sonic Mach number can $>10$ \citep{2006ApJ...653L.125P}. Very importantly, the high density features are typically perpendicular to the mean magnetic field, contrary to the common observations on the observed CNM emission lines \citep{2015PhRvL.115x1302C, 2020AnA...639A..26K}, suggesting that these shock-like features are unlikely the visible CNM in the HI emission lines but denser features in the interstellar media. A handy estimate of the aspect ratio of these features can be obtained by estimating how much compression can the dense features have in different thermal regimes: In the case of an adiabatic equation of state (diffuse part of CNM), $n=5/3$, and therefore the maximum compression is $\rho_{new}/\rho_{original} \sim 4$ . Assuming the gas was originally isotropic and being compressed by pressure, then the resultant aspect ratio is 4. For the denser part of CNM and diffuse $H_2$, the compression is at maximum $M_s$, and as a result, the aspect ratio is also $M_s$.

{\subsubsection{Super-Alfvenic Turbulence?}}
\label{superAlf}

While at first glance, super-Alfvenic turbulence does not appear to have any relation to magnetically aligned filaments, we have to remind the readers that the cold neutral media is believed to be mildly super-Alfvenic (See, e.g., \citealt{2018A&A...619A..58K}), despite the fact that the ambient warm and unstable phases are in the sub-Alfvenic regime. We will go through the complexity of multiphase turbulence later in this section, but we would like to first outline some basic properties of super-Alfvenic turbulence.

In the case of isothermal MHD turbulence and if $V_L > V_A$, at large scales, magnetic back-reaction is not important, and up to the scale (see \citealt{Lazarian06}):
\begin{equation}
l_A = L_{inj} M_A^{-3},
\label{la}
\end{equation}
the turbulent cascade is essentially a hydrodynamic Kolmogorov cascade. At the scale $l_A$, the turbulence transfers to the trans-Alfvenic turbulence described by GS95 scalings, i.e., anisotropy of turbulent eddies starts to occur at scales smaller than $l_A$, in a form very similar to Eq.\ref{eq:gs95_full} but without the $M_A^{-4/3}$ factor and replacing $L_{inj}$ with $l_A$ \citep{2019ApJ...878..157X}:
\begin{equation}
l_\parallel = l_A^{1/3} l_\perp^{2/3} = L_{inj}^{1/3} M_A^{-1} l_\perp^{2/3}
\label{eq:gs95_superalf}
\end{equation}
In other words, small-scale features in super-Alfvenic turbulence are still anisotropic (See, e.g., \citealt{2010ApJ...720..742K}), but the deviation of the Alfvenic Mach number from unity {\it suppresses} the anisotropy of filaments. A similar discussion on how frame transformation applies to Eq.\ref{eq:gs95_superalf} (c.f. Eq.\ref{eq:aspect_gf}) can be performed, but since $M_A > 1$, filaments will appear to be randomly oriented around the mean magnetic field direction after projection along the line of sight.

\begin{figure}
\label{fig:GA_early_fig}
\centering
\includegraphics[width=0.65\textwidth]{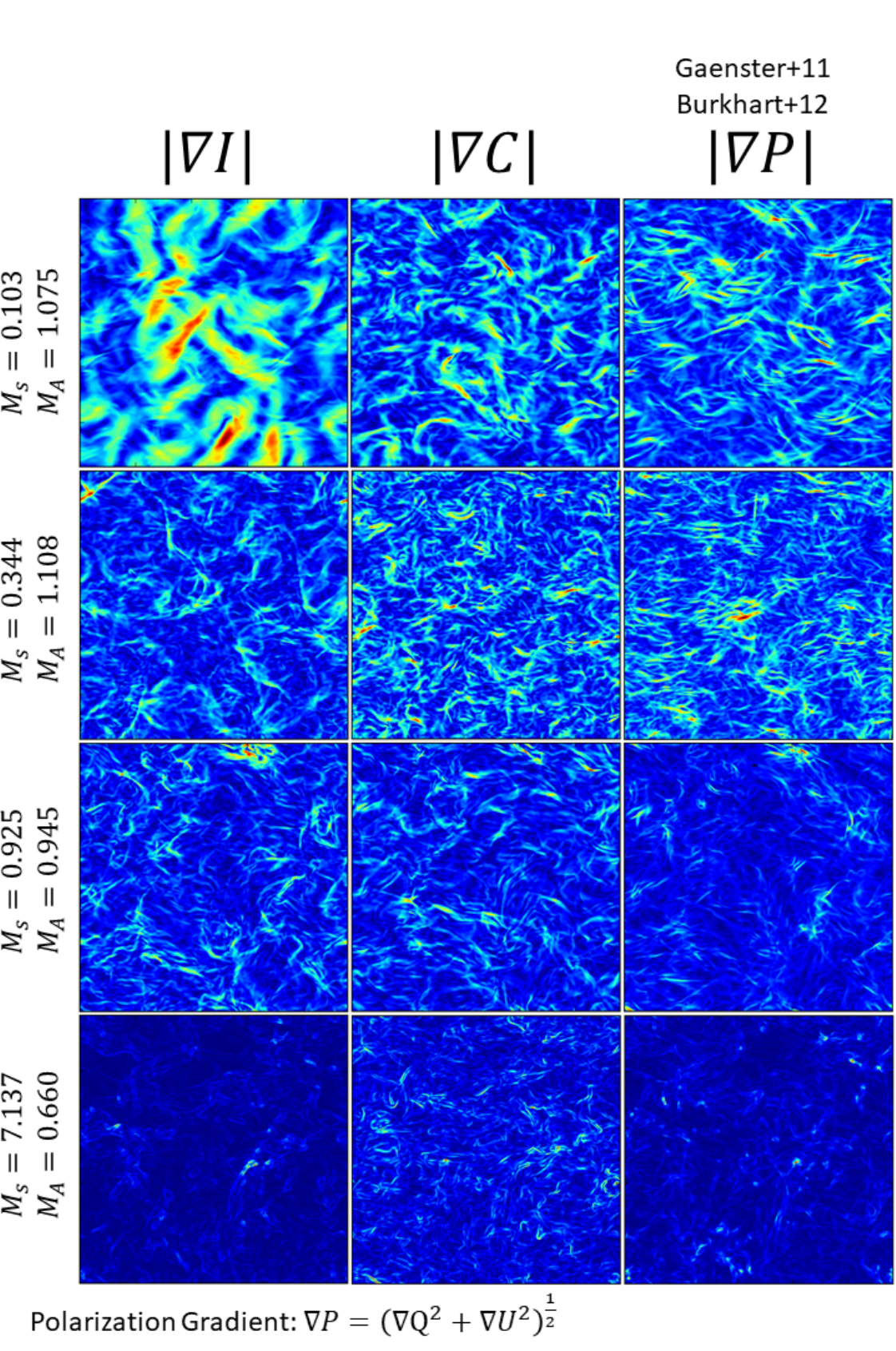}
\caption{A tabulation on how filamentary features visually look like for various $M_s$ and $M_A$. The mean field is parallel to the horizontal axis. From \cite{GA}. }
\end{figure}

\begin{table}
\centering
\begin{tabular}{c c c c}
Mode &  Power spectra $E(k)$ & Anisotropy factor & Frame vector \\ \hline \hline
Alfven & $k^{-5/3}$ & $\exp(-M_A^{-4/3} k_\parallel/k_\perp^{2/3}$) & $\zeta_A$\\
Slow (low $\beta$)  & $k^{-5/3}$ & varies\footnote{(see \cite{2020PhRvX..10c1021M})} & $\zeta_S$\\
Slow (high $\beta$) &  $k^{-5/3}$  & $\exp(-M_A^{-4/3} k_\parallel/k_\perp^{2/3})$ & $\zeta_S$\\
Fast (low $\beta$)  & $k^{-3/2}$ &  1 & $\zeta_F$\\
Fast (high $\beta$) & $k^{-3/2}$ & $k_\perp^2$ & $\zeta_F$ \\\hline\hline
\end{tabular}
\caption{\label{tab:theory_GS95}  A summary of the theoretical expectations of the turbulence scaling laws for sub-Alfvenic and subsonic turbulence. Summarized from \cite{CL03,2020PhRvX..10c1021M}. From \cite{leakage}.   }
\end{table}

\subsection{Multi-phase Micro-physics: Thermal instability, Turbulence-fueled unstable phases, Turbulence Heat Transport}

\subsubsection{Brief introduction}
Turbulence in the ISM is much more complicated. The ISM community usually divides the ISM study into two distinct categories: (1) the study of very diffuse, multiphase neutral HI turbulence, and (2) the study of ISM turbulence within molecular clouds that is typically supersonic, isothermal, trans- to sub-Alfvenic, and often sub-critical\footnote{This means the magnetic criticality, i.e., the ratio of gravitational energy to that of magnetic energy, is smaller than 1}. While early numerical works already explored the interaction of phase and turbulence \citep{2005A&A...433....1A,2007A&A...465..431H,2007A&A...465..445H}, it is only very recently (e.g., \citealt{2013A&A...556A.153H,2017NJPh...19f5003K,2016ApJ...819..137K,2017ApJ...834...25K,2017ApJ...846..133K,2018ApJ...853..173K,2018ApJ...859...68K,2020MNRAS.492.1465S,2020MNRAS.497.4196S,VDA,instability,2023ApJ...949L...5F}) that the community started studying how turbulence continuously cascades from the multiphase media to the dense molecular clouds.

The heating and cooling of ISM gases has been discussed multiple times in the literature (See, e.g., \citealt{2002ApJ...564L..97K,2003ApJ...587..278W,Draine2011,2016SAAS...43...85K}). Here, we discuss the basics and the consequences of the phase diagram depicted in Fig. \ref{fig:phase_diagram}. Table. \ref{tab:ISMtable} lists the phases of ISM and their typical physical parameters, and in Fig. \ref {fig:GA_early_fig} we show how varying $M_s$ and $M_A$ changes the shape of filaments in simulations. The most important heating process in the diffuse ISM is photoelectric heating. There are also multiple cooling processes. Ly-$\alpha$ dominates the cooling of the diffuse hot ISM, while $C^+$ dominates the cooling of the denser parts. Typically, one could model the cooling (of $C^+$) and heating via the following construction \citep{1983ApJ...270..511Z}:
\begin{equation}
\begin{aligned}
\text{\bf Heating rate}:& \Gamma_{pe} \propto n_{HI}\\
\text{\bf Cooling rate due to }C^+:& \Lambda \propto n_{HI}^2 \sqrt{T}\exp\left(-\frac{184}{T}\right)
\end{aligned}
\end{equation}
For the multiphase ISM, the typical equation of state (EOS) for HI is adiabatic for WNM and CNM: $P \propto \rho^{5/3}$. There is a particular range of densities (and temperatures) determined by the heating and cooling rates. Between the warm and the cold phase, there is a peculiar phase named {\bf Unstable Neutral Media} (UNM), which exhibits unusual physical properties related to turbulence. In particular, $P_{UNM}\propto \rho^n$ with $n<0$. The ubiquitous existence of UNM (about 20\% in both mass and volume fraction, \citealt{Kalberla2015}) is crucial for the formation stability of the CNM as turbulence interacts with UNM in a non-trivial manner. Readers should note that in some of the literature, there are exact cut-offs for the WNM and CNM temperatures. This is theoretically incorrect as the phase transition is a function of the two rates. The formation of the intermediate UNM phase is highly dependent on various factors. For instance, \cite{2003ApJ...587..278W} modeled the formation of UNM phases as a function of galactic distance. They found that at a distance of 18 kpc from the galactic center, the cooling rate is not sufficient for the formation of CNM. Turbulence in the WNM is subsonic and sub-Alfvenic, while for CNM it is supersonic \citep{2019A&A...627A.112K}. 
The turbulence conditions for the unstable phase are at the borderline and likely do not obey the critical balance condition (i.e., no GS95 scaling). Nevertheless turbulence cascades continuously over all three phases \citep{spectrum}.

In the case of molecular clouds, dust cooling and ambient HI-shielding play an important role in maintaining the isothermality of molecular clouds. Molecular clouds are typically modeled as isothermal, as temperature fluctuations are negligible (see Tab.\ref{tab:ISMtable}). Turbulence is typically highly supersonic ($M_s>10$) and highly compressible ($\beta\ll 1$). It is also well-known that molecular clouds are highly magnetized, either inferred from Zeeman measurements \citep{2010ApJ...725..466C} or the hourglass morphology observed in polarization \citep{li}. It is typical to model the turbulence in molecular clouds as trans-Alfvenic or sub-Alfvenic and highly self-gravitating, in contrast with the multiphase where self gravity has a minimal effect, see \citealt{2018MNRAS.476.4932V}).

\begin{table}[t]
\centering
\caption{\label{tab:ISMtable}The typical $M_s,M_A,\beta$ values for interstellar media and molecular clouds, $n_H$, $\delta v$ for molecular clouds from \cite{Draine2011,2018A&A...619A..58K,VDA,instability,cattail}, $\beta$ is from \cite{instability}.}
\begin{tabular}{c c c c c}
Phases & $n_H (cm^{-3})$ & $\delta v (km/s) $  & $\beta=2M_A^2/M_S^2$ & Typical temperature (K) \\
\hline
WNM                  & $0.1-1 $ & $10-17$ & $\sim 100$ & $5500-10000$\\
UNM                  & $1-10$   & $6-10$ & $\sim 1$ & $200-5500$\\
CNM                  & $10-50$  & $3-5$ & $\sim0.1$ & $50-200$\\
GMC Complex ($H_2$)  & $>10^2$   & $4-50$ & $\sim 0.01$ & $<50$  \\
\hline
\end{tabular}
\end{table}

\subsubsection{Thermal instability constrains cold filament anisotropy}

\begin{figure}
\label{fig:ho23_illustrative}
\centering
\includegraphics[width=0.99\textwidth]{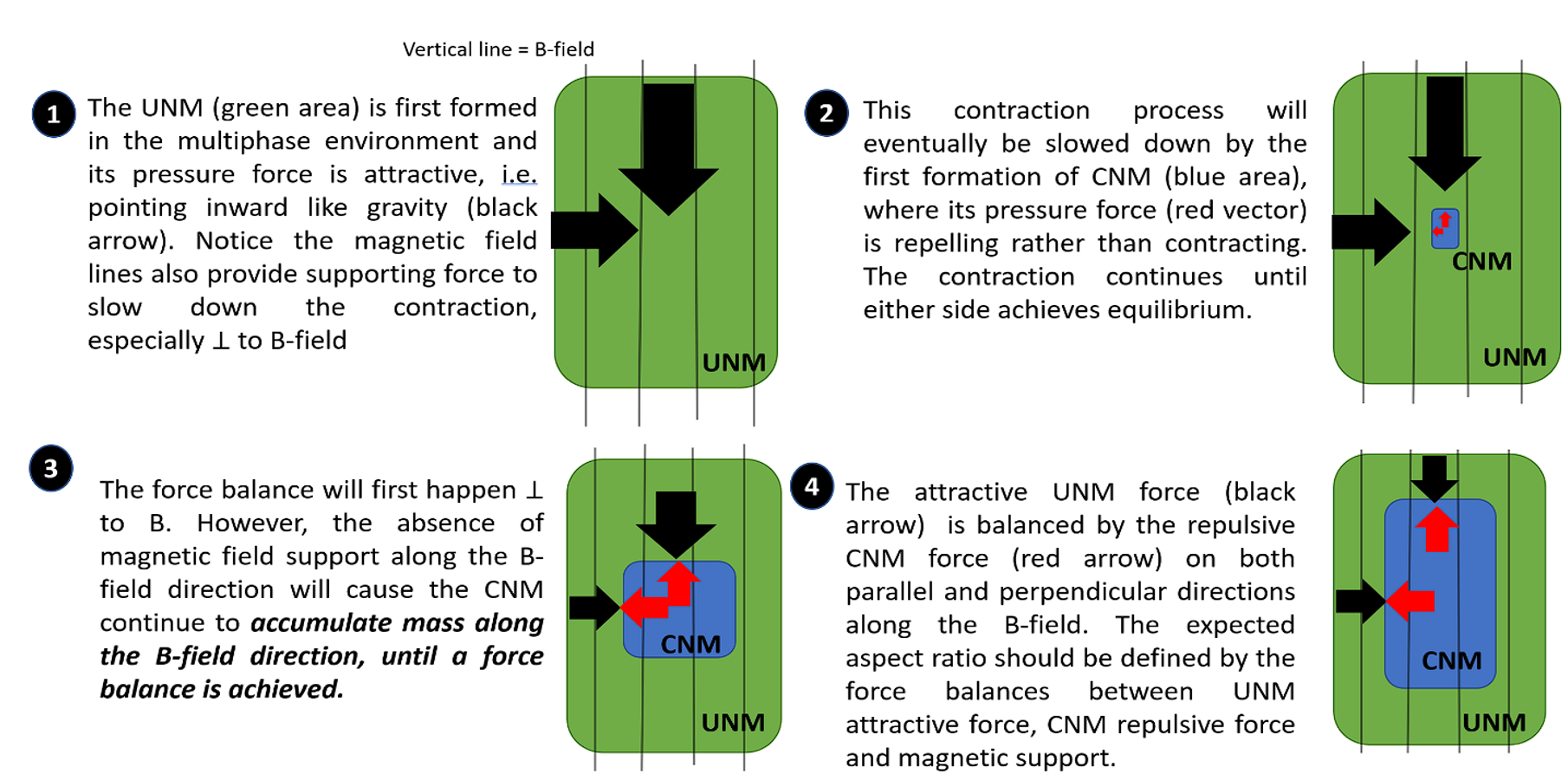}
\caption{An infographic describing the force balance model from \cite{instability} decides the length of CNM filament in magnetized multiphase system. Notice the aspect ratio of CNM predicted by this force balancing model is shorter than the GS95 estimate.}
\end{figure}

The existence of multiphase heating and cooling definitely affects the dynamics and spatial properties of the CNM because the assumptions of Goldreich-Sridhar turbulence (which holds at large scales) do not apply as we cascade from large roughly 10 pc to small scales roughly 1 pc. While the velocity perturbation $\delta v$ remains more or less constant \citep{instability}, both the thermal and Alfvenic speeds undergo significant changes during phase transitions. Consequently, the cold phase has both its sonic and Alfvenic Mach numbers one order of magnitude higher than those of the warm phase, even though $\delta v$ remains constant \citep{2018A&A...619A..58K}. In terms of the Goldreich-Sridhar critical balance postulate, this means that the phase speed of Alfven waves suddenly drops by a factor of 10, resulting in evident changes in the corresponding anisotropy formed under the Goldreich-Sridhar turbulence framework. Indeed, as observed by \cite{instability}, the actual measured filament aspect ratio from multiphase numerical simulations significantly deviates from the Goldreich-Sridhar estimate. This deviation appears to be consistent with earlier studies as well \citep{2013A&A...556A.153H}.

\begin{figure}
\label{fig:phase_diagram}
\centering
\includegraphics[width=0.47\textwidth]{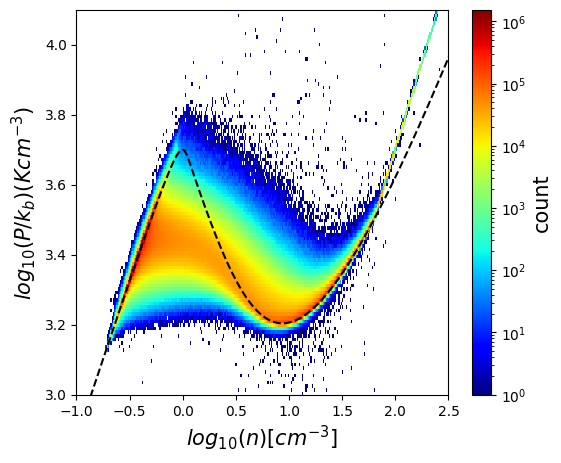}
\caption{A typical phase diagram for multiphase ISM simulations. From \cite{instability}. }
\end{figure}

\begin{figure*}
\label{fig:GS95_isothermal}
\centering
\includegraphics[width=0.47\textwidth]{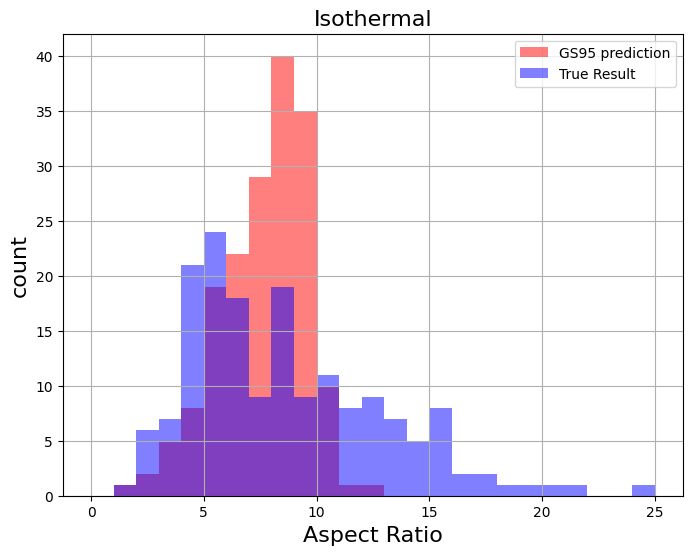}
\includegraphics[width=0.47\textwidth]{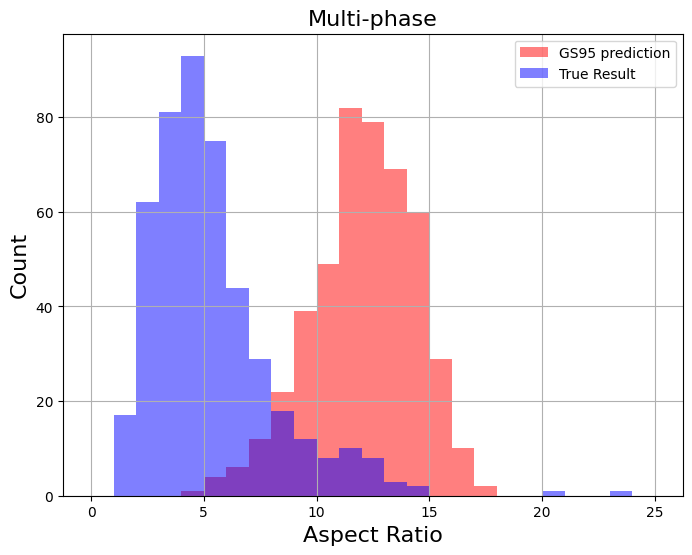}
\caption{Histogram of aspect ration between true filament aspect ratio and GS95 prediction ($\frac{l_\parallel}{l_\perp} \sim M_A^{-4/3} (\frac{L_{inj}}{l_\perp})^{1/3}$). Left: isothermal case ($M_S \approx 6, M_A \approx 0.8, N^3 = 792^3$. Right: Multi-phase strong {B-field} case.) Adopted from \cite{instability}.}
\end{figure*}

How do multiphase simulations (e.g., \citealt{2013A&A...556A.153H,2017NJPh...19f5003K,2018ApJ...853..173K,instability,2023ApJ...949L...5F}) differ from isothermal turbulence studies of \citep{CL03,2010ApJ...720..742K} in terms of turbulent cascades? Considering the "onion" model of the ISM \citep{MO07} and the prevalence and dominance of the UNM, \citep{2018ApJS..238...14M} in terms of both mass and volume fractions, it is likely that it plays a significant role in shaping the CNM. \cite{instability} pointed out that the presence of massive unstable phases acts as a stabilizer for the formed cold phase, and prevents the growth of log filaments. This is because maintaining a negative adiabatic index over a substantial period creates local pockets with inward-falling "gravity-like" pressure terms. These local pockets condense further and develop into cold clouds similar to the situation with Jeans instability. A graphic describing this process is given in Fig. \ref{fig:ho23_illustrative}. Any inward-falling collapsing forces are counterbalanced by turbulence and the magnetic field until the former becomes dominant, triggering instability. However, unlike gravity, the inward-falling force only acts within specific density and length scale roughly 1 pc ranges corresponding to the unstable phase.

\cite{instability} perform a quantitative analysis and recognize two regimes, depending on the relative pressure in the unstable and the cold phases (See Fig.\ref{fig:model}): 
\begin{enumerate}
\item { \bf UNM compression regime} 
\begin{equation}
0 < \frac{2}{25.6}(\frac{n_{peak}}{n_c}-1)<\frac{l_\bot}{l_{U,\bot}}.
\label{eq:aspect_UNM_compression}
\end{equation}

\item {\bf CNM expansion regime}
\begin{equation}
\frac{l_\parallel}{l_\bot} = \frac{M_S^2}{M_A} [ 2(\frac{n_{peak}}{n_C}-1) - 25.6  \frac{l_\bot}{l_{U,\bot}} ]^{-1}.
\label{eq:aspect_CNM_expansion}
\end{equation}
\end{enumerate}
which is very different from the GS95 expectation (See Fig.\ref{fig:GS95_isothermal}).

\begin{figure}
\label{fig:model}
\centering
\includegraphics[width=0.5\textwidth]{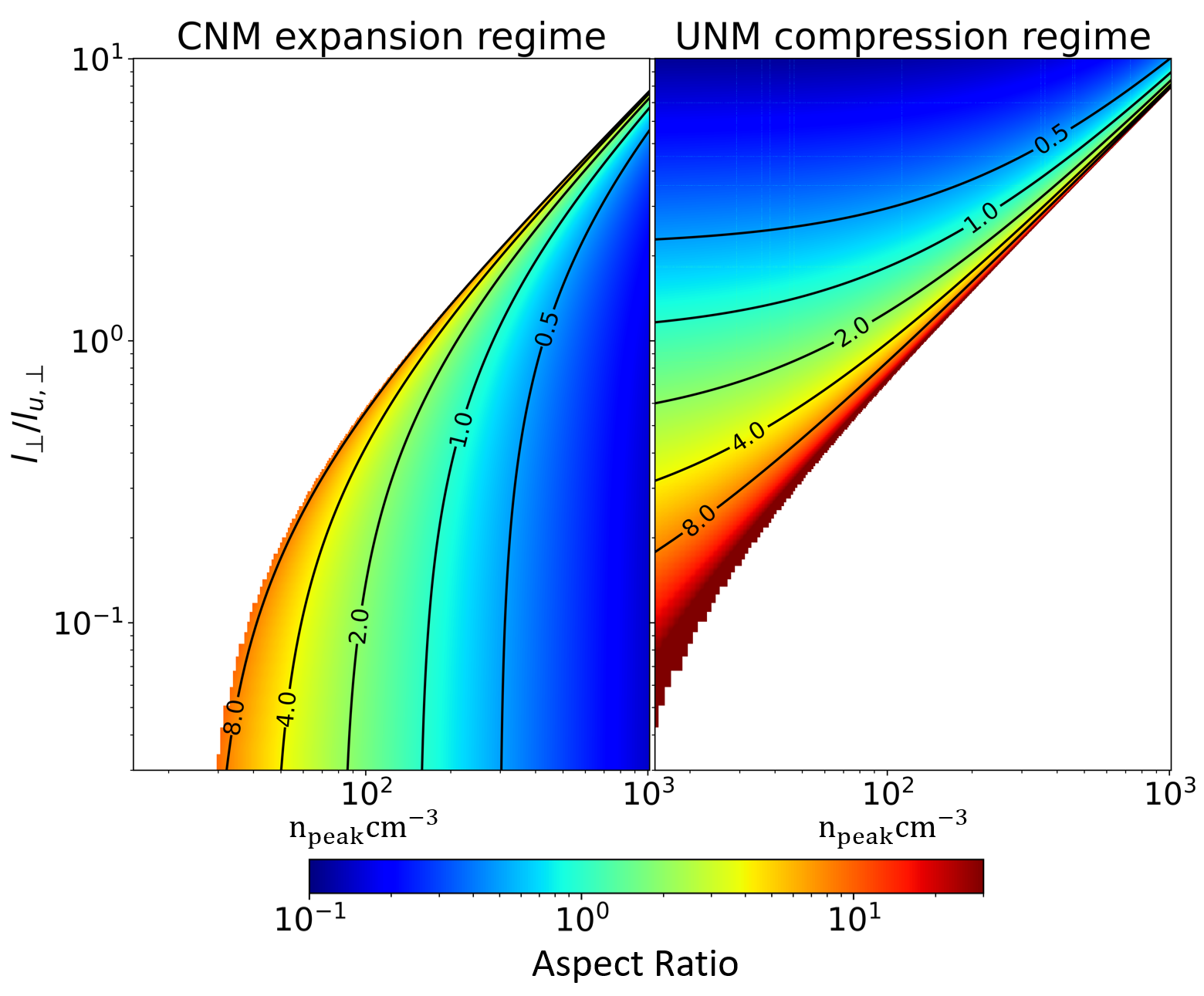}
\caption{The analytical prediction (Eq. \ref{eq:aspect_CNM_expansion}) illustrates how the aspect ratio varies as functions of the peak hydrogen number densities of the cold and unstable phases ($n_{\text{peak}}$) and the ratio of the widths of CNM and UNM (${l_{\bot}}/l_{U,\bot}$), using typical Mach numbers for CNM ($M_S \approx 5$) and $M_A \approx 0.8$. The iso-contours represent the values of the aspect ratio. When the aspect ratio is greater than 1, the filament is parallel to the magnetic field (B), while when it is less than 1, the filament is perpendicular to B. This analysis is adopted from \cite{instability}. }
\end{figure}

\cite{instability} observed that the filaments formed in the case of a multi-phase turbulence system generally have a shorter aspect ratio {\avi than that predicted by GS95}. Recall from the previous section that the maximum compression for adiabatic gases is 4. The unstable phase allows the condensation of the cold phase to reach a certain density threshold, but the aspect ratio of CNM formed in realistic multiphase numerical simulation remains less than that estimated using Goldreich-Sridhar. 
This is because the inward-falling pressure formed by phase dynamics is isotropic in nature. Long filaments (in the sense of greater than $\sim 20$) are dynamically unstable as long as the unstable phase is present for a sufficiently long period of time {\avi (See Figure \ref{fig:phi_gradient})}.

\subsubsection{Unstable phase fraction bumped up by turbulence}

Earlier numerical studies (e.g., \citealt{2013A&A...556A.153H,2017NJPh...19f5003K,2018ApJ...853..173K}) and observational studies (e.g., \citealt{2018ApJS..238...14M,2020AnA...639A..26K}) already show that the fraction of the unstable phase is significant. A natural question regarding the dynamics of the unstable phase, as discussed in the previous section, is whether the unstable phase should indeed be "unstable" and dissipate rather quickly with a short dynamical time, as proposed in \citealt{1977ApJ...218..148M}. An important question natrually arises: how does the interstellar medium sustain such a large amount of what should be unstable phases?

Earlier literature (e.g., \citealt{2001ApJ...557L.121G,2002ApJ...569L.127K,2003ApJ...587..278W}) already suggested that turbulent motions play a crucial role in determining the temporal and spatial structure of the phases. Of particular interest is the regime discussed in \citealt{2003ApJ...587..278W}, where $\rho(P_{WNM,max})<\rho<\rho(P_{CNM,min})$, leading to the natural appearance of two stable phases. Physically, when turbulence levels are strong enough, more cold phases are cycled back to the warmer phases. This phenomenon has been observed when turbulence levels are sufficiently high \citep{2017NJPh...19f5003K}. However, the exact mechanisms by which turbulence controls the three-phase structure remain a mystery.

Figure \ref{fig:phi_gradient} illustrates how variations in turbulence energy injection affect the relative fraction of unstable phases. Clearly, as more turbulent energy is injected into the turbulence box, the fraction of the unstable phase increases. Moreover, the phases remain in a stable fraction for an extended period, and with higher turbulent energy injection, the cold phase enters a steady state more rapidly. This may appear counter-intuitive at first glance. However, when we consider the "stability" effect discussed in the previous section, it becomes evident that the increased fraction of the longer-lasting unstable phase actually contributes to the formation and maintenance of both the temporal and spatial stability of cold gases.
\begin{figure*}
\label{fig:phi_gradient}
\includegraphics[width=0.48\paperheight]{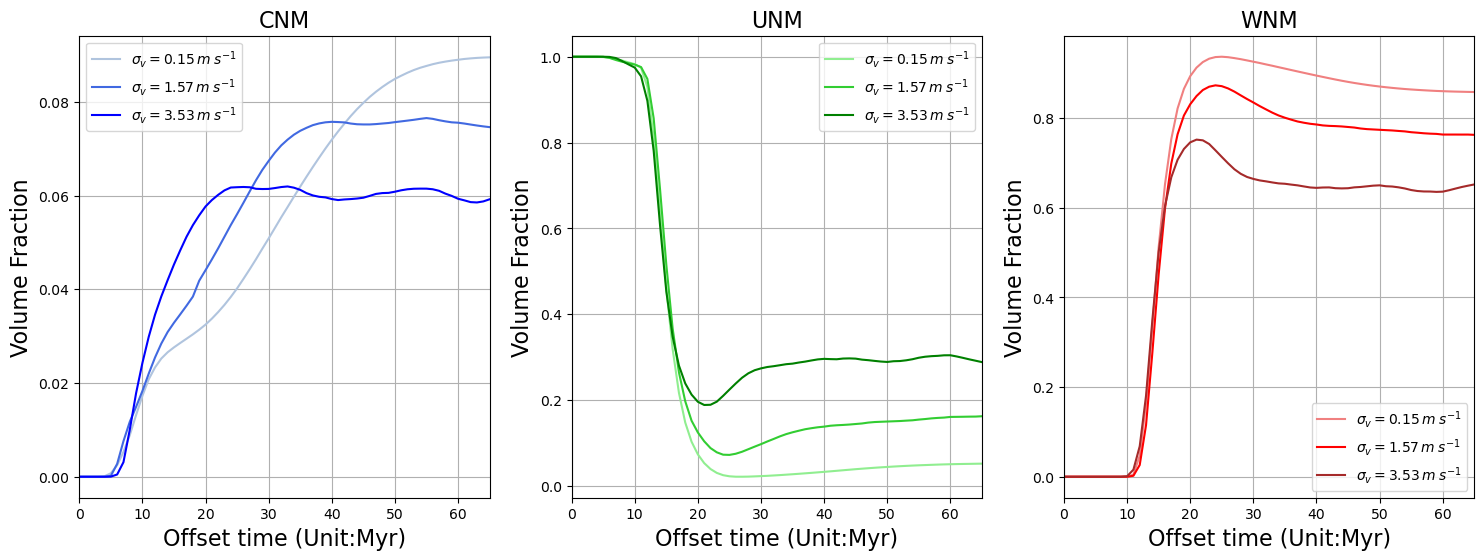}
\caption{The evolution of three phases in three multiphase simulations with different turbulence levels. It is rather clear that (i) the increase of turbulent velocity dispersion (denoted by $\sigma_v$) increase the fraction of unstable phases (ii) the three phases are in dynamical equilibrium for an extended period of time (> 20-40Myr depending on the phases). From Ho, Yuen, Lazarian (submitted). }
\end{figure*}

\subsubsection{Heat transport from magnetized multiphase turbulence}

In the previous two sub-subsections, we discussed the unexpectedly long lifetime of the unstable phase and how  it prevents the formation of long filaments. However, the relationship between heat transfer and turbulence remains unresolved. Naively, turbulence acts as both a free energy source that pumps up the cold phase and sustains the thermal phase in the unstable regime. The latter forces the cold phases to condense. In other words, turbulence acts as both a heat source and a stabilizer for the unstable-cold phase complex, creating some sort of equilibrium between them.

Turbulence can also diffuse energy in a similar fashion to thermal diffusion. In the case of the more diffuse intra-cluster medium \citep{2001ApJ...562L.129N,2012ApJ...754..122K,2020ApJ...889L...1L}, the heat transfer due to thermal conduction and turbulent transport is of comparable magnitude (only differing by a factor of 5 compared to Spitzer's \citep{1962pfig.book.....S} value when the magnetic field is chaotic \citep{2001ApJ...562L.129N}). \cite{2003ApJ...589L..77C} briefly discussed that in the case of magnetic field wandering and mixing (See \citealt{LV99,2012NPGeo..19..297K}) the heat transport by turbulence is about 100 times more effective than thermal conduction. However, \cite{2003ApJ...589L..77C} used fictional temperature values for the cold and warm phases that significantly deviate from modern observations (see Tab.\ref{tab:ISMtable}). Correcting for this fact would yield a ratio of "turbulent conductivity" to Spitzer's value \citep{1962pfig.book.....S} on the order of $\kappa_{turb}/\kappa_{Sp} \sim O(1000)$, suggesting that heat exchange is mainly performed by turbulence transport. In the cold phase, the existence of turbulence quickly equalizes the temperature within it.

Does anisotropic turbulence provide anisotropic temperature equalization within cold neutral media? Thermal instability from anisotropic conduction was proposed as early as \cite{1965ApJ...142..531F} and later extensively studied by \cite{2012ApJ...747...86C}, where the latter assumes an anisotropic heat conduction tensor. A characteristic length from \cite{1965ApJ...142..531F} suggests that heat can be equalized within one "Field's length":
\begin{equation}
\lambda_F = \frac{\text{Conduction-Diffusion}}{\text{Cooling}} \propto \left(\frac{\kappa T}{\rho^2 \Lambda}\right)^{1/2}
\label{eq:fields}
\end{equation}
which determines the "coherent thermal length" of a cold filament. \cite{2012ApJ...747...86C} construct an anisotropic thermal tensor and propose that the parallel conductivity $\kappa_\parallel \propto \langle B\rangle^2$, while that of the perpendicular conductivity $\kappa_\perp \propto \delta B^2$, which gives an aspect ratio scaling to Alfvenic Mach number ($\delta B/\langle B\rangle \sim M_A$):
\begin{equation}
\frac{\lambda_{F,\parallel}}{\lambda_{F,\perp}} \sim M_A^{-1}
\end{equation}
which they test in numerical simulations of the intra-cluster medium. However, this construction via anisotropic thermal tensor has an immediate issue if we want to apply this argument to the ISM: Thermal transport in the ISM, as we argued earlier, is dominated by turbulence.

One obvious method of estimating the effect of turbulence thermal transport via the Field's length argument is to replace $\kappa$ in Eq. \ref{eq:fields} with $\kappa_{\text{dynamic}}$, which can be approximated using the characteristic length and velocity from a given turbulence model:
\begin{equation}
\kappa_{\text{dynamic},\parallel/\perp} \rightarrow l_{\parallel/\perp} v_{\text{turb},\parallel/\perp}
\end{equation}
This estimation yields a scaling of $\frac{\lambda_{F,\parallel}}{\lambda_{F,\perp}} \propto M_A^{-4/3}$, similar to the direct GS95 estimate performed in the previous subsection (Eq. \ref{eq:ar_gs95}). However, this direct substitution does not consider the effect of turbulence-thermal phase interaction that we discussed in the previous two sub-subsections.

So, how do we model the aspect ratio of cold neutral media with the knowledge from the previous subsections? Yuen et al. (submitted) performed a quantitative treatment of how turbulence interacts with radiative heating and CII cooling, changing the heat transport properties in the multiphase ISM. It was discovered that the adiabatic index, which controls Eqs. \ref{eq:aspect_UNM_compression} and \ref{eq:aspect_CNM_expansion}, is related to the plasma Mach numbers: $|n| \propto M_s^2/M_A^2$. From there, Yuen et al. (submitted) discussed that the anisotropy can be larger than the thermal instability estimate of the above equations by flattening $n$, while still staying smaller than the GS95 estimate (Eq. \ref{eq:ar_gs95}).

\subsection{Observational Constraints: Projection, Mapping, Velocity caustics and absorption}

Observational constraints can hinder the application of the above theories.How 3D turbulent information is mapped into 2D observational space was comprehensively developed by the framework of \cite{LP00} (see also \citealt{LP04,LP06,LP08,LP12,LP16,KLP16,KLP17a,VDA,leakage}). The essence of the LP statistical theory aims to answer a very simple question: For a given statistical model of turbulence in 3D, how do the statistics of observables behave? While this question may sound trivial, the difficulty increases dramatically when magnetic fields and turbulence are both in effect, not to mention the effects of radiative transfer (in particular, absorption). There are two types of transfer considered under the framework of LP statistical theory:
\begin{enumerate}
    \item How is the observed column density or velocity-like observables correlated with the 3D statistics of turbulent density and velocity in emissions \citep{LP00,KLP16,KLP17a,VDA,GA,curvature,ch5} and absorption maps \citep{LP04,LP06,LP08}?
    \item How are the observed polarization fluctuations (e.g., Stokes Q, U, polarization angle, polarization percentage) correlated with the 3D fluctuations of polarization (\citealt{LP12,LP16,ch5,leakage,ch9,2023MNRAS.524.6102M})?
\end{enumerate}
These questions are exceptionally critical when discussing "filaments" observed via different instruments. A simple example is that for a given 3D filament oriented at a certain nontrivial angle with respect to the line of sight, the projection of it will appear as a shorter filament. A less trivial examples are the filaments observed in the velocity channel space. When HI absorption is in play (Theory: \citealt{LP04,LP06,LP08,2016A&A...595A..32B}, Numerics: \cite{2022MNRAS.512.4765S}, Observation: \citealt{2018ApJS..238...14M}), a significant portion of the cold phases is actually optically thick. Are the filaments that we observe in HI emission maps cold or unstable phases? In this subsection, we will briefly go through the theories of statistical mapping \citep{LP00,LP04,LP06,LP08,LP12,LP16,KLP16,KLP17a,VDA,leakage}, and then we will discuss how they impact our statistical measurements of filaments when using different methods outlined in Section 2.

\subsubsection{Statistical Mapping: Intensity and Centroid Map}

The construction of the statistical theory for observables usually involves correlation and structure functions using a suitable frame. Since MHD turbulence exhibits anisotropy along the mean magnetic field, the statistical properties of the 3D turbulence variables differ parallel and perpendicular to the mean magnetic field. Earlier literature \citep{1946RSPSA.186..480B,1950RSPTA.242..557C,1981PhRvA..24.2135M,1997PhRvE..56.2875O} has utilized this property to characterize turbulence anisotropy. We can represent the correlation function (and also the structure function) of a vector turbulence variable $X_{i=x,y,z}$ in the axis-symmetric form:
\begin{equation}
\langle {\bf X}_{i}({\bf r'}){\bf X}_{j}({\bf r}+{\bf r'})\rangle_{\bf r'} = A\delta_{ij} + B\hat{r}_i\hat{r}_j + C\hat{\lambda}_i\hat{\lambda}_j + D(\hat{r}_i\hat{\lambda}_j+\hat{\lambda}_i\hat{r}_j)
\label{eq:XXR}
\end{equation}
Here, $\langle ...\rangle_{\bf r'}$ denotes volume averaging over the 3D position vector ${\bf r'}$, and $\hat{\lambda}=\hat{B}$ is the mean field unit vector. Note that $A,B,C,D$ are all functions of ${\bf r}$. Due to axis-symmetry, it suffices to write ${\bf r} = (r,\mu = \cos^{-1}(\hat{r}\cdot\hat{\lambda}))$. Similar expressions can be done in Fourier space with $\hat{k}$ and $\hat{\lambda}$.

The axis-symmetric model can be further expressed using frame vectors. Mathematically, in the case when the frame vectors are orthogonal, denoted as $(\hat{\zeta}^{A},\hat{\zeta}^{F},\hat{\zeta}^{P})$ (see Tab.\ref{tab:notations}), the traceless\footnote{i.e., removing the A-term, which occurs when the variable itself has a zero mean} tensor of $\langle {\bf \tilde{X}}{i}({\bf k'}){\bf \tilde{X}}{j}({\bf k}+{\bf k'})\rangle_{\bf k'}$ in the Fourier space can be represented in a linear, separable form:
\begin{equation}
\langle {\bf \tilde{X}}_{i}({\bf k'}){\bf \tilde{X}}_{j}({\bf k}+{\bf k'})\rangle_{\bf k'} = C_A \hat{\zeta}^{A}\hat{\zeta}^{A} + C_F\hat{\zeta}^{F}\hat{\zeta}^{F} + C_P\hat{\zeta}^{P}\hat{\zeta}^{P}
\label{eq:XXK}
\end{equation}
Here, $C_{A,F,P}$ are constants. Notably, for the case of $X$ representing magnetic field fluctuations or velocity fluctuations in incompressible turbulence, $C_P=0$, resulting in a highly simplified expression that is axis-symmetric.

When we observe Eq. \ref{eq:XXR} or Eq. \ref{eq:XXK} in observations, an additional anisotropy is introduced depending on how the line of sight is oriented relative to the frame vectors $(\hat{\zeta}^{A},\hat{\zeta}^{F},\hat{\zeta}^{P})$ \citep{leakage}. Earlier studies \citep{2020NatAs...4.1001Z} utilized this fact to quantify mode signatures in synchrotron polarization maps, while more recently \citep{2023MNRAS.524.6102M} used the statistical relations to determine the line-of-sight angle. In \cite{KLP16} and later \cite{KLP17a}, they discussed how the anisotropy of both intensity and velocity centroid correlation functions (Eq. \ref{eq:XXR}) is related to (a) mode fraction, (b) the underlying turbulence model, and (c) the inclination angle of the mean magnetic field relative to the line of sight. The full context of \cite{KLP16,KLP17a} is beyond the scope of this article; however, we want to emphasize that the geometric relations are not simple linear or monotonic functions of the three dependencies (a-c) we listed above (see also \cite{ch5} for more examples).

\subsubsection{Velocity Channel map and velocity caustics}
\begin{figure*}[th]
  \centering
  \includegraphics[width=0.99\textwidth]{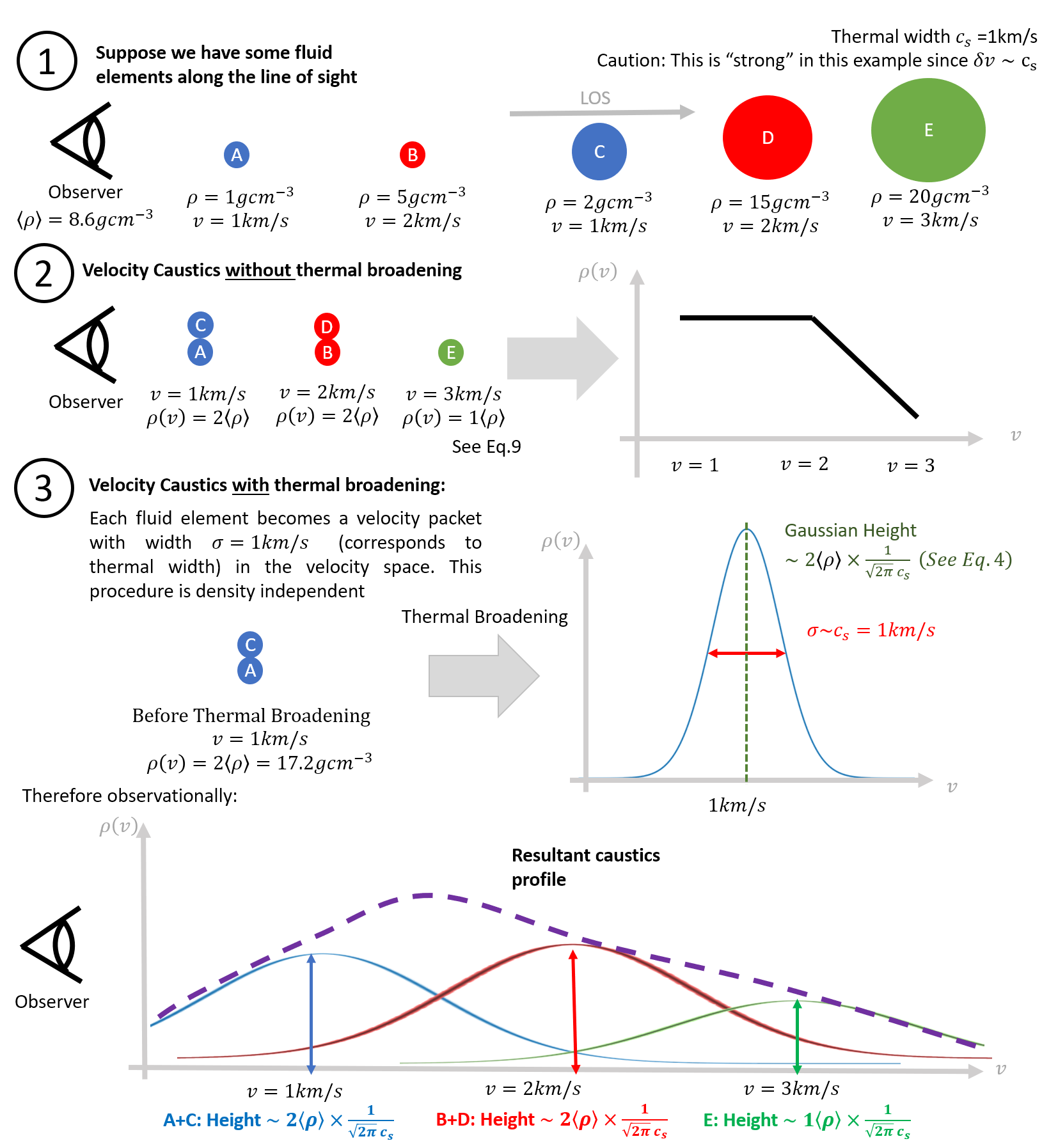}
  \caption{\label{fig:cartoons} A cartoon illustrating how velocity caustics are formed in a spectroscopic position-position-velocity (PPV) cube and the concept of velocity and density fluctuations in PPV data. From the top: Panel (1) shows our example that facilitates the discussion of velocity caustics. Panel (2): We show how the velocity caustics, based on the example in (1), look like without thermal broadening. Panel (3): The velocity caustics with thermal broadening, in the view of the $\rho(v)-v$ diagram, where we plot the caustics profile (the purple dashed line) as a function of $v$. From \cite{VDA}. }
\end{figure*}

\begin{figure*}[th]
  \centering
  \includegraphics[width=0.99\textwidth]{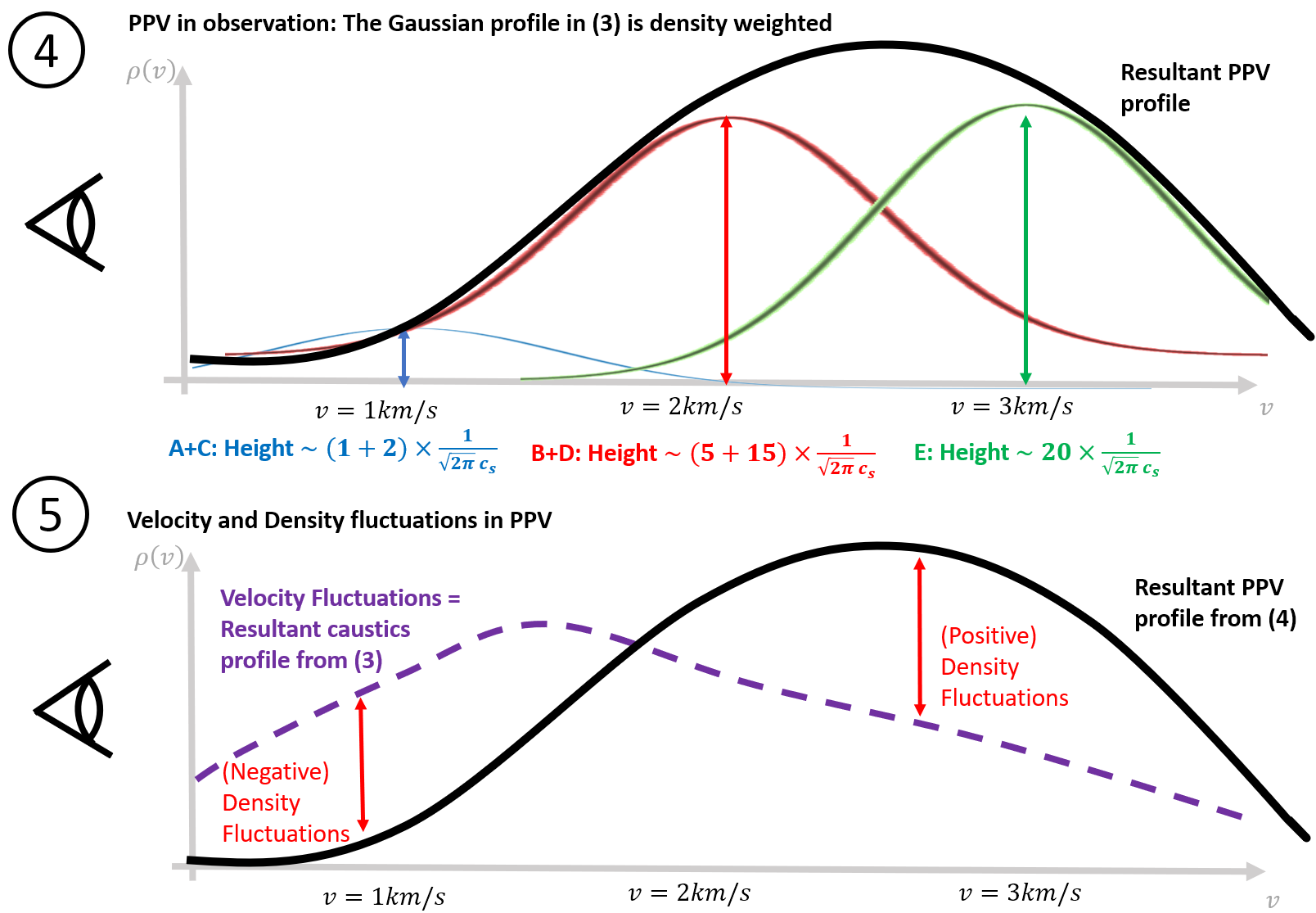}
  \caption{{\bf (continuing Fig.\ref{fig:cartoons})}. Panel (4): The real PPV profile in observations (the black line) according to our example in panel (1). Panel (5): The differences between the caustics profile and the true PPV profile correspond to the density fluctuation, while the caustics profile itself represents the velocity fluctuation. From \cite{VDA}.}
\end{figure*}

\begin{figure}[th]
  \centering
  \includegraphics[width=0.49\textwidth]{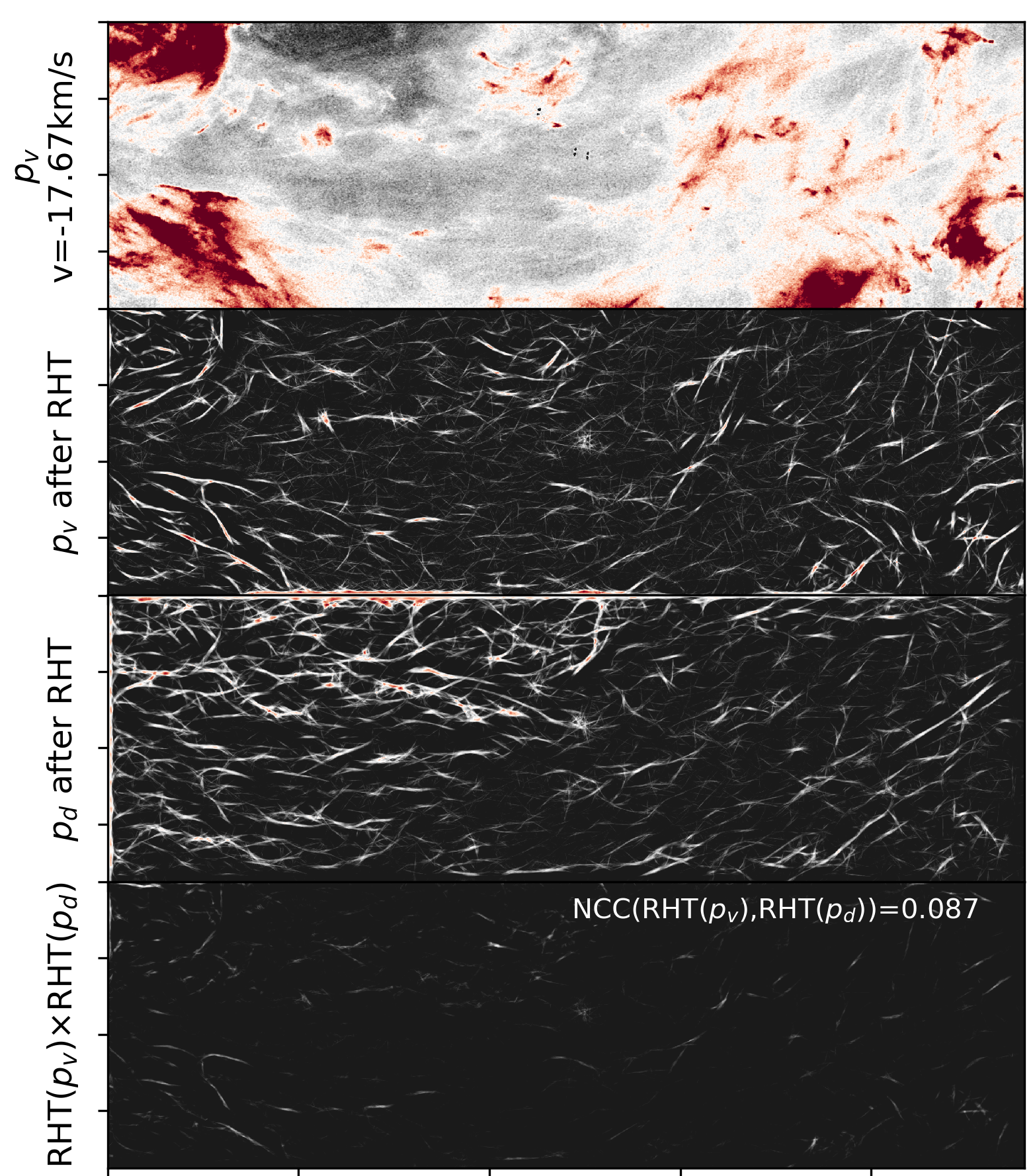}
  \includegraphics[width=0.49\textwidth]{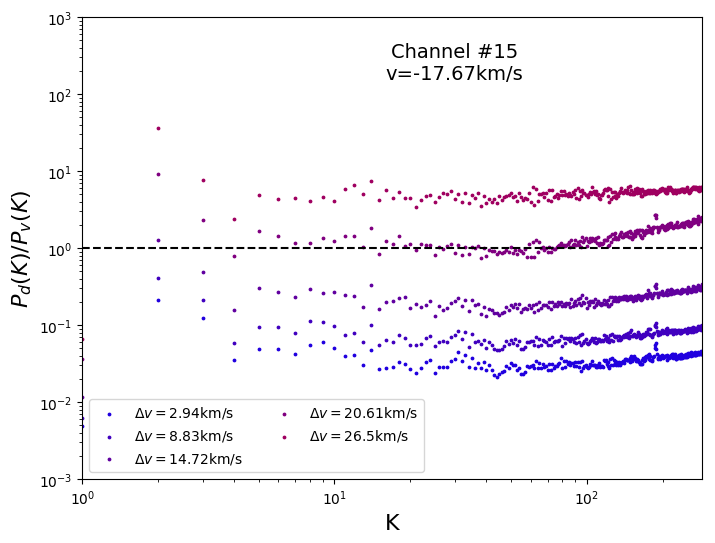}
  \caption{\label{fig:RHT} Left panel: A set of figures showing how a selected wing channel ($v=-17.67 km/s$, $\Delta v = 2.94 km/s$) from the region in \cite{2015PhRvL.115x1302C} would look like after Velocity Decomposition Algorithm (VDA,\citealt{VDA}), and the output of RHT from VDA maps. From the top panel: The $p_v$ map after VDA. 2nd and 3rd panels from the top: The RHT results of $p_v$ and $p_d$, respectively, scaled to $[0,1]$. The lower panel: The product of the RHT results of $p_d$ and $p_v$, scaled to $[0,1]$. The normalized correlation coefficient ($NCC(A,B) = \frac{\langle (A-\langle A\rangle)(B-\langle B\rangle)\rangle}{\sigma_A\sigma_B}$, see \cite{2019arXiv190403173Y}, $NCC\in[-1,1]$) is 0.087, which means the RHT results of $p_d$ and $p_v$ are basically uncorrelated. Right panel: The spectra ratio $P_d/P_v$ at the wing channel from the HI data from \cite{2015PhRvL.115x1302C} at $v=-17.67 km/s$ and $\Delta v = 2.94 km/s$. We can see that as the channel width goes smaller, the spectra ratio quickly drops below one for most $K$. From \cite{VDA}.
 }
\end{figure}

Perhaps the most important observation quantity for HI emission map is the velocity channel map.  In \cite{LP00} it was analytically described how the density and velocity fluctuations contribute to the fluctuation of velocity channels. The density in PPV space of emitters with temperature $T$ moving along the line-of-sight with stochastic turbulent velocity $v_{turb}(\bf x)$ and regular coherent velocity $v_{\mathrm{g}}(\bf x)$ is (See \citealt{LP04}):
\begin{equation}
\begin{aligned}
p(\mathbf{X},v_0,\Delta v) &= \int dz \rho(\mathbf{X},z) \left(\frac{m}{2\pi k_BT}\right)^{1/2} \times\\
&\int_{v_0-\Delta v/2}^{v_0+\Delta v/2} dv W(v) e^{-\frac{m(v-v_{turb}(\mathbf{X},z))^2}{2k_{B}T(\mathbf{X},z)}},
\label{eq:rho_PPV}
\end{aligned}
\end{equation}
where sky position is described by 2D vector $\mathbf{X}=(x,y)$ and $z$ is the line-of-sight coordinate and $W(v)$ is a window function given by the instrument. Formally we can always write the fluctuation of velocity channel intensity as:
\begin{equation}
  p(\mathbf{X},v_0,\Delta v) - \langle p\rangle_{{\bf X}\in A} = p_d(\mathbf{X},v_0,\Delta v) + p_v(\mathbf{X},v_0,\Delta v)
  \label{eq:sp}
\end{equation}
where $\langle p\rangle_{{\bf X}\in A}$ represents the velocity channel averaged over a certain spatial area $A$. The subtraction of the mean value in Eq.\ref{eq:sp} is required as we deal with the fluctuations arising from turbulence. Notice that $p_d$ and $p_v$ are functions of the Plane of Sky (POS) two dimensional vector $\mathbf{X}$, as well as the velocity channel position $v_0$ and channel width $\Delta v$. In what follows, we shall refer to $p_v$, i.e. the velocity contribution to velocity channels, as the {\bf velocity caustics contribution}. It is noted by \cite{LP00} that in the case of thin ($\Delta v \ll 1$) channel map is dominated by velocity fluctuations. A cartoon from \cite{VDA} summarizes the science behind the velocity caustics fluctuations (Fig.\ref{fig:cartoons}).

There was a proposal \citep{Clark19} that utilized a $128^3$ isothermal simulation and a statistically unnormalized parameter (See criticism from \citealt{2019arXiv190403173Y}) to claim that the fluctuations from channel maps arise from density fluctuations from cold neutral media, which was also taken for granted by \cite{2022arXiv220201610K}. It was revealed by a systematic study by \cite{VDA} that the velocity and density fluctuations can be separated in channel maps and they can have a varying fraction depending on the turbulence properties.

Summarized by \cite{VDA}, there are several properties that the density ($p_d$) and velocity ($p_v$) contributions obey:
\begin{enumerate}
  \item { Orthogonality of $p_d$ and $p_v$ when $M_s\ll 1$}
  \item { $p_v=0$ when $\Delta v\rightarrow \infty$}
  \item { $p_d\propto I$ when $M_s \ll 1$}
\end{enumerate}

With these items, \cite{VDA} {\avi created the Velocity Decomposision Algorithim (VDA) which} derived analytically the formula for $p_v$ an $p_d$ which enables the extraction of density and velocity fluctuations in channel maps:
\begin{equation}
\begin{aligned}
  p_v &= p - \left( \langle pI\rangle-\langle p\rangle\langle I \rangle\right)\frac{I-\langle I\rangle}{\sigma_I^2}\\
  p_d &= p-p_v\\
  &=\left( \langle pI\rangle-\langle p\rangle\langle I \rangle\right)\frac{I-\langle I\rangle}{\sigma_I^2}
\end{aligned}
\label{eq:ld2}
\end{equation}

An observational application of Eq.\ref{eq:ld2} and the statistical analysis is shown in Fig.\ref{fig:RHT}. It is evident from Fig.\ref{fig:RHT} that:
(i) The velocity caustics contribution is never zero and is spatially uncorrelated to density features.
(ii) The relative contribution of density and velocity contribution is a function of channel width, particularly in the thin channel case where it is the fluctuations of velocity that give rise to filamentary features in the channel map.

What is the impact of \cite{VDA} for the study of filaments in HI emission maps? It is evident that if the channel map is velocity-like (i.e. $\langle p_v^2\rangle>\langle p_d^2\rangle$ in Eq.\ref{eq:ld2}), then the GS95 of filaments (c.f. \citealt{2019ApJ...878..157X}, Eq.\ref{eq:ar_gs95},\ref{eq:ar_gs952}) can be directly applied without consideration of complicated density processes, as we discussed in Section 3.1.2. Furthermore, by varying the channel width, we can obtain both velocity and density statistics of filaments, allowing us to directly compare them with the multiphase physics as outlined in Section 3.2.

\subsubsection{HI Absorption, and HI-H2 Chemistry}
\begin{figure*}[th]
  \centering
  \includegraphics[width=0.99\textwidth]{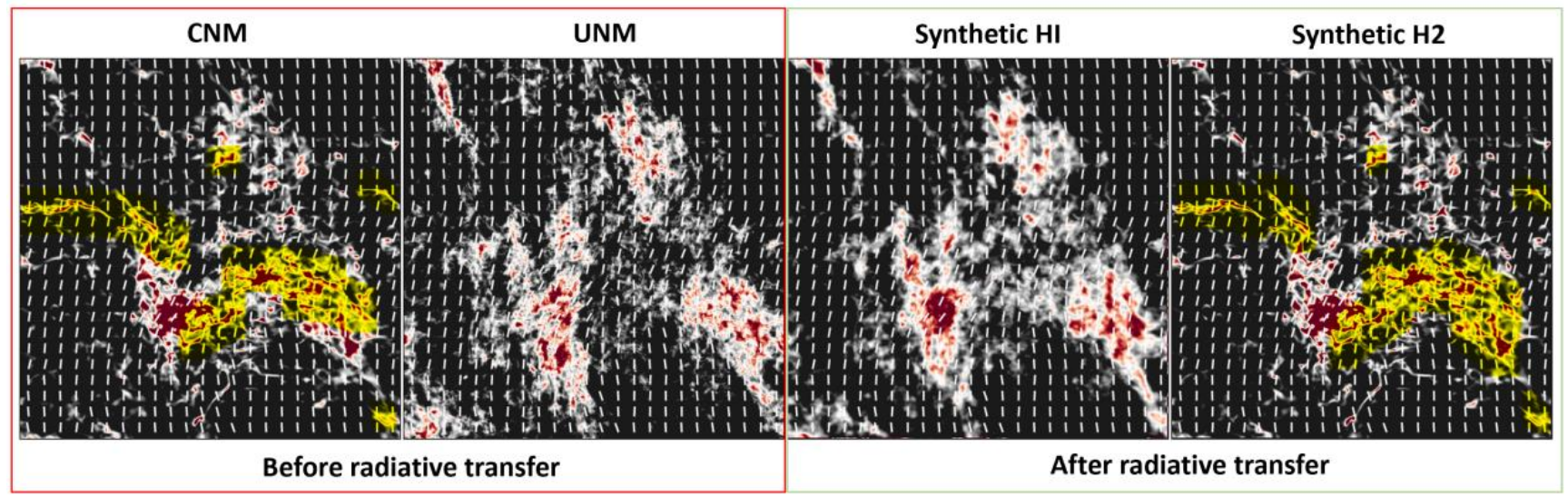}
  \caption{\label{fig:synthetic}  (Left 2 panels) A set of figures shows how the actual distributions of CNM (far left) and UNM (second left) column densities look like in our multiphase simulation before radiative transfer after 20 Myr has passed in simulations (right two panels). We perform an observational synthesis and produce the synthetic HI column densities (second right) and the H2 column densities (far right) overlaid with the projected B-field (white vector). One can see that both CNM and synthetic H2 have a significant portion of features being perpendicular to the B-field (yellow region). However, for UNM and synthetic HI map, they have almost zero perpendicular features. Modified from \cite{VDA}.}
\end{figure*}

The effect of absorption \citep{2020MNRAS.497.4196S} and HI-H2 conversion \citep{2016ApJ...822...83B,2016A&A...595A..32B,2017ApJ...835..126B,2023arXiv230504965G} creates extra complications in deciphering how the emission map of HI looks and how to interpret the filamentary features therein. In this section, we provide a brief highlight of how these two effects contribute to the observational appearance of HI.

As discussed in Section 3.1, features with high density, low $\beta$, and a high sonic Mach number preferentially align perpendicular to the magnetic field. This alignment is well observed in simulations of both isothermal (e.g., \citealt{2010ApJ...720..742K,LY18a}) and multiphase media (\citealt{2018PhRvL.121b1104K,VDA,instability}). However, these high-density features are essentially unseen in observational HI emission maps \citep{2014ApJ...789...82C,2015PhRvL.115x1302C,2017A&A...607A..15K,2018A&A...619A..58K}. To address this issue, two primary mechanisms lower the emission strength of ultra-dense ($>\sim 50 cm^{-3}$) cold neutral media: the high optical depth of cold neutral media and the sharp $HI-H_2$ conversion density threshold, causing these ultra-dense features to convert to H2 and become dark at 21cm lines. \cite{2020MNRAS.492.1465S} discusses how self-absorption of HI favors the presence of field-parallel filaments in observational maps. 

An additional effect is the HI-H2 conversion, where \cite{2016ApJ...822...83B} showed that the transition from HI to H2 occurs rapidly, with one species dropping by an order of magnitude in density when the density increases. We can model this transition via radiative transfer processes by incorporating arguments related to both optical depth and HI-H2 conversion. Fig. \ref{fig:synthetic} illustrates the synthetic results due to these optical effects. From this figure, it becomes evident that most of the features perpendicular to the magnetic field are extinguished in synthetic HI emission maps due to the combined effects of HI-H2 transition and high optical depth.


\section{Summary and conclusions}
\label{sec5}
In Section 1, we posed two key questions:
\begin{enumerate}
\item {\bf Why are observed CNM filaments only parallel to the magnetic field while simulations contain both parallel and perpendicular orientations?} 

Filaments form naturally in MHD turbulence, regardless of the phases. In the case of isothermal supersonic low $\beta$ MHD turbulence (Section 3.1), there are typically two types of density enhancements: parallel (to magnetic field) filaments which have lower density, and perpendicular filaments which have higher density. These filaments are relatively stable in isothermal turbulence (in timescales of eddy over time) but thermal instabilities cause long filaments to fragment.

Thermal instabilities occur when the presence of radiative heating and cooling creates a nontrivial equation of state (Section 3.2, i.e. $P\propto \rho^n$ with $n<0$.), in which the pressure term at a certain range of densities becomes gravity-like instead of supportive. As a result, long perpendicular filaments are fragmented into smaller pieces in timescales shorter than the eddy turnover time. Multiphase MHD turbulence simulation shows that perpendicular filaments that have densities nearing the cold phase are thermally unstable. As a result, the perpendicular filaments are less extended on the sky {\it provided that the gas tracers could see the entire density ranges of multiphase interstellar turbulence}.

Observational constraints of 21 cm lines further complicates the CNM seen in inferometric measurements (Section 3.3). CNM becomes optically thick and preferentially perpendicular to magnetic fields right at its own density threshold. As a result, 21 cm measurements traces only the optical thin, long filamentary features that are preferentially parallel to magnetic field, but the perpendicular filaments are self-absorbed. Furthermore, the HI-H2 conversion becomes very effective right after the formation of CNM. The perpendicular filaments are therefore mostly converted to molecular hydrogen, which are dark in 21 cm lines. 

The channel map effect (Section 3.3) also creates additional impact to the relative ratio of parallel and perpendicular filaments seen in observations. Velocity crowding effect in the presence of turbulence generates filamentary features in channel map, despite no real physical density enhancements. Unlike density fluctuations, turbulent velocity fluctuations are mostly parallel to magnetic field even in the case of supersonic limit. As a result, the amount of parallel filaments observed in 21cm lines are significantly increased due to velocity crowding. The effect of velocity crowding can be diminished by the Velocity Decomposition Algorithm.

\item  {\bf     Why does observed HI have an aspect ratio of hundreds while simulations do not?}\linebreak
As previously explained the long aspect ratio observed in HI filaments are likely attributed to algorithmic effects, projection effects, and the conflation of velocity and density fluctuations in H1 channel maps. 

The filament detection method (Section 2) together with the velocity crowding effect (Section 3.3) will create illusory aspect ratios of filaments. Filaments that are next to each other, oriented in the same way and classified with edge detection method are likely be identified as a thin filament with high aspect ratio (Section 3.1). As we have outlined both algorithmically (Section 2) and analytically (Section 3.1, 3.3), observed 21 cm lines are subjected to numerous effects along the line of sight, on top of the true physical effects from turbulence and thermal instabilities. Determining whether the filamentary structures observed in channel map or intensity maps are really true filaments, sub-structures from a larger system, or artifacts is necessary before seeking physical explanation of why filaments carry certain aspect ratios.

Filaments with large aspect ratios are generally not thermally stable. Turbulence plays a crucial role in extending and stabilizing the spatial and temporal properties of these fragmented filaments. The extended lifetime of unstable neutral medium (UNM), due to its lower Alfvenic mach number and its role as a thermal cushion, is essential for explaining why CNM does not rapidly dissipate by its supersonic turbulence. CNM, while not gravitationally bound, can persist stably before fully converting into H2. The strong correlation between a high fraction of UNM and turbulence indicates that UNM fraction could serve as a turbulent diagnostic for statistical studies. Additional support by the heat transport effect due to turbulence anisotropy is explored in Section 3.2.4. However, how exactly the interplay between turbulence and heat transport impacts the aspect ratio of physical filaments is still a subject of research.

\end{enumerate}

Despite significant advancements in observational and simulation capabilities, many mysteries surrounding the interstellar medium and its role in star formation persist. Ongoing efforts, such as large-scale simulations and observations, hold promise for shedding light on these questions and providing a deeper understanding of the interstellar medium and its influence on star formation processes. Notable endeavors in this direction include large-scale simulations \citep{2018PhRvL.121b1104K,2018ApJ...853..173K,2019A&A...622A.166S,2023ApJ...949L...5F} and observations \citep{2018ApJS..234....2P,Kalberla2015,2018ApJS..238...14M,2022ApJ...926..190R,2022ApJ...928...79R,2022ApJ...926..186D}, which aim to uncover more precise properties of the interstellar medium and investigate its subsequent impact on star formation processes.

\appendix	
\section{Notations and abbreviations}
. 
\linebreak\linebreak
\begin{table}[h]
\centering
\small
\caption{\label{tab:notations}List of notations used in this review.}
\begin{tabular}{l c c}  
\hline\hline
{\bf Parameter} & {\bf Meaning}\\
\hline\hline
${\bf r}$ & 3-D separation ${\bf x}_2-{\bf x}_1$ \\ 
${\bf R}$ & 2-D separation ${\bf X}_2-{\bf X}_1$\\ 
$z$ & Line of sight (LOS) variable \\ \hline
${\bf x}$ & 3-D position vector\\
${\bf X}$ & 2-D position vector\\
$l$ & Distance of the 3d separation \\ 
${\cal L}$ & Size of a turbulent cloud\\
${L_{inj}}$ & Turbulence injection scale\\ \hline
$A$ & A(lfven)-type vector component\\
$F$ & F-type vector componentl $=C$-type \\
$P$ & P(otential)-type vector component\\ \hline
$\rho({\bf r})$ & 3-D Density\\ 
$\rho({\bf X},v)$ & Emitters' intensity in the PPV space \\
$B$ & 3-D magnetic field\\ 
$b_{turb}$ & Turbulent part of the magnetic field\\
$B_\perp$ & Projected magnetic field\\
$B_{x,y}$ & The x \& y components of the magnetic field\\
$Q,U$ & Stokes Q \& U parameters\\
$v$ & 3-D velocity\\ 
$C$ & Velocity Centroid\\ 
$\theta$ & Magnetic field angle\\
$\phi$ & Polarization angle\\ \hline
$M_s$ & Sonic Mach number\\ 
$M_A$ & Alfvenic Mach number\\ 
$n$ & Adiabatic index\\
 \hline
$\langle A \rangle_{x}$ & average of the quantity $A$ over variable $x$\\
$\gamma$ & Angle between the line of sight and symmetry axis\\
$\mu$ & $=\widehat{k}\cdot \widehat{B}$, $\mu=cos(\gamma)$\\
O(n) & Big-O notation\\
\hline \hline
\end{tabular} 
\end{table}

\begin{table}[t]
\centering
\small
\caption{\label{tab:abbreviations} List of abbreviations used in this review, in alphabetical order}
\begin{tabular}{l c c}  
\hline\hline
{\bf Abbreviations} & {\bf Full name}\\
\hline\hline
CNM & Cold Neutral Media\\
GS95 & \cite{GS95}\\
MHD & Magneto-hydrodynamics \\
$H_2$ & Molecualar Hydrogen\\
HI & Neutral Hydrogen\\
HII & Ionized Hydrogen\\
LP & Lazarian \& Pogosyan series (See \cite{LP00})\\
RHT & Rolling Hough Transform (See \citealt{2015PhRvL.115x1302C})\\
SNR & Signal to Noise\\
UNM & Unstable Neutral Media\\
USM & Unsharp Mask  \\
VDA & Velocity Decomposition Algorithm (See \citealt{VDA})\\
VGT & Velocity Gradient Technique (See \citealt{YL17a})\\
WNM & Warm Neutral Media\\
\hline \hline
\end{tabular} 
\end{table}

%

\noindent{\bf Acknowledgment} Inspiring discussions with Edouard Audit, Alex Lazarian, Alexei Kritsuk, and Chris McKee are acknowledged. Research presented in this article was supported by LANL LDRD-20220700PRD1. We also thank our collaborators who made much of the science discussed in this paper possible. KHY would like to thank the AAPPS-DPP committee for the invitation to deliver a talk on the subject and to write this review paper.. \linebreak

{\noindent{\bf Conflict of Interest}} On behalf of all authors, the corresponding author states that there is no conflict of interest.

\bibliography{refs}

\end{document}